\documentclass[pra,floatfix,preprint,nofootinbib,superscriptaddress,eqsecnum,12pt]{revtex4}
\usepackage{epsfig}
\usepackage{bm}
\usepackage{amsmath}
\input epsf
\numberwithin{equation}{section}

\begin{document}
\setlength{\baselineskip}{.2in}

\title{TIME SYMMETRY AND ASYMMETRY IN\\
           QUANTUM MECHANICS AND QUANTUM COSMOLOGY}

\author{Murray Gell-Mann}
\affiliation{Lauritsen Laboratory\\
 California Institute of Technology, Pasadena, CA 91125}

\author{James B. Hartle}
\email{hartle@physics.ucsb.edu}
\affiliation{Department of Physics\\
 University of California, Santa Barbara, CA 93106-9530}

\date{\today}

\maketitle

\section{Introduction}

The disparity between the time symmetry of the fundamental laws of
physics and the time asymmetries of the observed universe has been a
subject of fascination for physicists since the late 19th 
century.\footnote{For clear reviews, see \cite{Dav76, Pen79, Zeh89}.}
The following general time asymmetries are observed in this universe:

\begin{itemize}

\item The thermodynamic arrow of time --- the fact that
approximately isolated systems are now almost all evolving towards equilibrium
in the same direction of time.

\item The psychological arrow of time --- we remember the
past, we predict the future.

\item The arrow of time of retarded electromagnetic
radiation.

\item The arrow of time supplied by the $CP$ non-invariance of
the weak interactions and the $CPT$ invariance of field theory.

\item The arrow of time of the approximately uniform 
expansion of the universe.

\item The arrow of time supplied by the growth of 
inhomogeneity in the expanding universe.

\end{itemize}

All of these time asymmetries could arise from time-symmetric dynamical
laws solved with time-asymmetric boundary conditions.  The thermodynamic
arrow of time, for example, is implied by an initial condition in
which the progenitors of today's approximately isolated systems were all far from
equilibrium at an initial time.  The $CP$ arrow of time could arise as a
spontaneously broken symmetry of the Hamiltonian. The
approximate uniform expansion of the universe and the growth of
inhomogeneity 
follow from an initial ``big bang'' of sufficient spatial homogeneity
and isotropy, given the attractive nature of gravity.  
Characteristically such arrows of time 
can be reversed temporarily, locally, in isolated subsystems,
although typically at an expense so great that the experiment can
be carried out only in our imaginations.  If we could, in the
classical example of Loschmidt \cite{Los1876}, 
reverse the momenta of all particles and
fields of an isolated subsystem, it would ``run backwards'' with
thermodynamic and electromagnetic arrows of time reversed.

Quantum cosmology is that part of physics concerned with the theory of
the boundary conditions of our universe.
It is, therefore, the natural and most general context in which to
investigate the origin of observed time asymmetries.  In the context
of contemporary quantum cosmology, several such
investigations have been carried 
out \cite{Pen79, Pag85, Haw85, HHaw85, Lafup, Haw87, HP88, Wad90},
 starting with those of Penrose \cite{Pen79} 
on classical time-asymmetric initial
and final conditions and those of Page \cite{Pag85} and
Hawking \cite{Haw85} on the emergence of the thermodynamic 
arrow of time from the
``no-boundary'' theory of the  initial condition of the universe.  It is not
our purpose to review these results or the status of our understanding
of the time asymmetries mentioned above.  Rather, we shall
discuss in this essay, from the perspective of quantum cosmology, 
a time asymmetry not
specifically mentioned above.  That is the arrow of time of familiar quantum
mechanics.

Conventional formulations of quantum mechanics incorporate a fundamental
distinction between the future and the past, as we shall review in
Section II.  This quantum-mechanical arrow of time has, in a way, 
a distinct status
in the theory from the time asymmetries discussed above.  It does not
arise, as they do, from a time-asymmetric choice of boundary conditions
for time-neutral dynamical laws.  Rather, it can be regarded as a time 
asymmetry of the
laws themselves.  However, the quantum mechanics of cosmology
does not have to be formulated in this time-asymmetric way.  In Section
III, extending discussions of Aharonov, Bergman, and Lebovitz \cite{ABL64}
and of Griffiths \cite{Gri84}, we consider a generalized quantum mechanics for
cosmology that utilizes both initial and final conditions to give a
time-neutral, two-boundary formulation that does not necessarily have an
arrow of time \cite{Har91}.  In such a formulation all time asymmetries arise
from properties of the initial and final conditions, in particular
differences between them, or,  
at particular epochs, from nearness to the
beginning or end.  A theory of both initial and final conditions would
be the objective of quantum cosmology.

In the context of a time-neutral formulation, the usual quantum
mechanics results from utilizing a special initial
condition, together with what amounts to a final condition representing complete
indifference with respect to the future states, thus yielding the
quantum-mechanical arrow of time, which is 
sufficient to explain the observed time
asymmetries of this universe. However, a time-neutral formulation of
quantum mechanics allows us to investigate to what extent the familiar
final condition of indifference with respect to future states is mandated
by our observations.  In particular, it allows us to investigate whether
quantum cosmologies with less blatantly asymmetric initial and final
conditions might also be consistent with the observed general time
asymmetries.  As a step in this direction we discuss 
a quantum cosmology that would be, in a sense, the
opposite extreme --- a cosmology with a time-symmetric pair of initial
and final conditions leading to a universe that is statistically
symmetric about a moment of time.  Such boundary conditions imply
deviations from the thermodynamic arrow of time and the arrow of time
supplied by the $CP$ non-invariance of the weak interactions.
We investigate such deviations to see
if they are inconsistent with observations.  The
classical statistical models reviewed in Section IV and the models of $CP$
symmetry breaking discussed in Section V suggest that the predicted
deviations may be insufficient to exclude time-symmetric boundary
conditions
if the interval between initial and final conditions
is much longer than our distance in time from the initial condition. 
Next, we review and augment the arguments of Davies and Twamley that 
electromagnetic radiation may supply a probe of the final condition
that {\it is} 
sufficiently accurate to rule out time-symmetric boundary conditions.

We should emphasize that we are not
advocating a time-symmetric cosmology but only using it as a foil to
test the extent to which observation now requires the usual asymmetric
boundary conditions and to search for more refined experimental tests.
 The important result of this paper is a quantum
framework for examining cosmologies with less asymmetric boundary
conditions than the usual ones, so that the quantum-mechanical arrow of
time (with its consequent time asymmetries) can be treated, or derived,
as one possibility out of many, to be confronted with observation,
rather than as an axiom of theory.

Relations between the initial and final conditions of a 
 quantum-mechanical universe {\it sufficient} for both $CPT$-symmetric cosmologies and
time-symmetric cosmologies are discussed in Section V.  Ways in which
the $T$-violation exhibited by the weak interaction  could arise in 
such universes
are described there as well.  In Section VI we discuss the limitations
on time-symmetric quantum boundary conditions following from the
requirements of decoherence and classicality.  Specifically, we show that
for a set of alternative histories to have the
negligible interference between its individual members that is
necessary for them to be assigned probabilities at all, there must be
some impurity in the initial or final density matrices or both, except
in the highly unorthodox case in which there are only one or two
coarse-grained histories with non-negligible probability.

We should make clear that our discussion of time-symmetric cosmologies,
based on speculative generalizations of quantum mechanics and
causality,  with
separate initial and final density matrices that are related by time
symmetry, 
 is essentially different from the conjecture
that  has sometimes been made that ordinary causal quantum or classical
mechanics, with just a single boundary condition or a single prescribed
wave function, $CPT$-invariant about some time in the distant future,   
might lead to a $T$-symmetric
or $CPT$-symmetric cosmology with a contracting phase in which the arrows
of time are reversed \cite{Gol62, Gelup67, Ne'70, Haw85, Zeh93}. 
It is the 
latter notion, by the
way, that Hawking refers to as his ``greatest mistake'' \cite{Haw93}. We shall
return to this topic in Section V. 

\section{The Arrow of Time in Quantum Mechanics}

As usually formulated, the laws of quantum mechanics are not time-neutral
but incorporate an arrow of time.  
This can be seen
clearly from the expression for the probabilities of histories
consisting of
alternatives at definite moments of time $t_1<t_2<\cdots < t_n$. 
Let $\{\alpha_k\}$ be an exhaustive set of alternatives at
time $t_k$ represented by $\{P_{\alpha_k}^k(t_k)\}$, a set of projection operators in
the Heisenberg picture. For example, the alternatives $\{\alpha_k\}$
might be defined by an exhaustive set of ranges for the center-of-mass
position of a massive body.  A particular history corresponds to a
specific sequence of alternatives $(\alpha_1, \cdots, \alpha_n)$.  The
probability for a particular history in the exhaustive set of histories is 
\begin{equation}
p\left(\alpha_n,\cdots, \alpha_1\right) =
Tr\Bigl[P^n_{\alpha_n}(t_n)\cdots P^1_{\alpha_1}(t_1)\rho
P^1_{\alpha_1}(t_1)\cdots
P^n_{\alpha_n}(t_n)\Bigr]\, ,
\label{twoone}
\end{equation}
where $\rho$ is the density matrix describing the initial state of the
system and the projection operators are time-ordered from the density
matrix to the trace.\footnote{This 
compact expression of the probabilities of ordinary quantum mechanics
has been noted by many authors.  For more details of this and other
aspects of the quantum mechanical formalism we shall employ the reader
is referred to \cite{GH90a} and \cite{Har91} where references to earlier literature may
be found.}

The expression for the probabilities \eqref{twoone} is not time-neutral.  This
is not because of the time ordering of the projection
operators.  Field theory is invariant under $CPT$ and the ordering of the
operators could be reversed by a $CPT$ transformation of the projection
operators and density matrix, leaving the probabilities unchanged.  (See
{\it e.g.} \cite{GH90a} or \cite{Har91}).  Either time ordering may therefore be used;  it is by
convention that we usually use the one with the condition represented  by
the density matrix $\rho$ in the past.

Rather, \eqref{twoone} is not time-neutral because there
is a density matrix on one end of the chain of projections representing
a history while at the other end there is the trace \cite{ABL64, Gri84, Har91}.
Whatever conventions are used for time ordering, there is thus an asymmetry
between future and past exhibited in the formula for probabilities
\eqref{twoone}.  That asymmetry is the arrow of time in quantum mechanics.

The asymmetry between past and future exhibited by quantum mechanics
implies the familiar notion of causality.  From an effective density matrix
describing the present {\it alone} it is possible to predict the probabilities for the future.  
More precisely, given that alternatives $\alpha_1, \cdots, \alpha_k$ have ``happened'' at times
$t_1 <\cdots < t_k$ before time $t$, the conditional probability
for alternatives $\alpha_{k+1}, \cdots, \alpha_n$ to occur in the future
at times $t_{k+1}, \cdots, t_n$ may be determined from an effective
density matrix $\rho_{\rm eff} (t)$ at time $t$.  Specifically, the 
conditional probabilities for future prediction are
\begin{equation}
p(\alpha_n, \cdots, \alpha_{k+1} | \alpha_k, \cdots, \alpha_1) =
\frac{p(\alpha_n, \cdots, \alpha_1)}{p(\alpha_k, \cdots, \alpha_1)}\, .
\label{twotwo}
\end{equation}
These can be expressed as
\begin{equation}
p(\alpha_n, \cdots, \alpha_{k+1} | \alpha_k, \cdots, \alpha_1) = Tr
\left[P^n_{\alpha_n} (t_n) \cdots P^{k+1}_{\alpha_{k+1}} (t_{k+1})
\rho_{\rm eff} (t_k) P^{k+1}_{\alpha_{k+1}} (t_1) \cdots P^n_{\alpha_n}
(t_n)\right]\, ,
\label{twothree}
\end{equation}
where the effective density matrix $\rho_{\rm eff}$ is
\begin{equation}
\rho_{\rm eff} (t_k) = \frac{P^k_{\alpha_k} (t_k) \cdots P^1_{\alpha_1}
(t_1) \rho P^1_{\alpha_1} (t_1) \cdots P^k_{\alpha_k}
(t_k)}{Tr\left[P^k_{\alpha_k} (t_k) \cdots P^1_{\alpha_1} (t_1) \rho
P^1_{\alpha_1} (t_1) \cdots P^k_{\alpha_k} (t_k)\right]}\, .
\label{twofour}
\end{equation}
The density matrix $\rho_{\rm eff} (t_k)$ can be said to define the
effective state of the universe at time $t_k$, given the history
$(\alpha_1, \cdots, \alpha_k)$.

What is the physical origin of the time asymmetry in the basic laws of
quantum mechanics and what is its connection with the other observed time
asymmetries of our universe?  The rest of this Section addresses that
question.

The reader may be most familiar with the expression for probabilities
\eqref{twoone} in the context of the approximate ``Copenhagen'' quantum mechanics
of measured subsystems.  In that case operators, the density matrix,
etc.~all refer to the Hilbert space of the subsystem.  The sets of
projection operators $\{P^k_{\alpha_k}(t_k)\}$ describe alternative
outcomes of measurements of the subsystem.  

Formula \eqref{twoone} for the probabilities of a sequence of
measured outcomes is then a unified expression of the ``two forms of
evolution'' usually discussed in the quantum mechanics of subsystems  --- unitary evolution
in between measurements and the ``reduction of the state vector'' on 
measurement.  The time
asymmetry of \eqref{twoone} does not arise from the unitary evolution of the
projection operators representing the measured quantities in the
Heisenberg picture;  that is time-reversible.  Rather,
it can be said to arise from the successive reductions represented by the
projections in \eqref{twofour} that occur on measurement.  The common explanation
for the origin of the arrow of time in the quantum mechanics of measured
subsystems is that
measurement is an irreversible process and that quantum mechanics
inherits its arrow of time from the arrow of time of
thermodynamics.\footnote{This connection between the thermodynamic arrow
of time and the quantum-mechanical arrow of time can be ambiguous.
Suppose, for example, a measuring apparatus is constructed in which the
local approach to equilibrium is in the opposite direction of time from
that generally prevailing in the larger universe.  If that apparatus
interacts with a subsystem (perhaps previously measured by other
apparatus adhering to the general thermodynamic arrow of time) should
the operators representing those measurements be ordered according to
the thermodynamic arrow of the apparatus or of the larger universe with
respect to which it is running backwards?  Such puzzles are resolvable
in the more general quantum mechanics of closed systems to be discussed
below, where ``measurements'', the ``thermodynamic arrow of time'', and any
connection between the two are all approximate notions holding in only
special situations.}  If that is the case, then the origin of the
quantum-mechanical arrow of time must ultimately be cosmological, for the
straightforward explanation of the thermodynamic arrow of time is a
special initial condition for the universe implying that its
constituents were far from equilibrium across a spacelike surface.  
Let us, therefore, investigate more fundamentally the quantum-mechanical 
arrow of time, not in an approximate quantum mechanics of
measured subsystems, but in the quantum mechanics of a closed system ---
most realistically and generally the universe as a whole.

The formula \eqref{twoone} for the probabilities of histories also holds in the
quantum mechanics of a closed system such as the universe as a whole, at
least in an approximation in which gross fluctuations in the geometry of
spacetime are neglected.  The sets of projection operators describe
alternatives for the whole system, say the universe,
 and the density matrix can be thought of as describing its  
initial condition.\footnote{For a more
detailed exposition of this quantum mechanics of cosmology, the reader is
referred to our previous work \cite{GH90a, GH90b, Har91}, where references to
the earlier literature may also be found.}  Not every set of histories
that may be described can be assigned probabilities according to eqref{twoone}.
In the quantum mechanics of closed systems consistent probabilities
given by \eqref{twoone} are predicted only for
those sets of histories for which there is negligible
interference between the individual members of the set \cite{Har91} as a consequence of the
particular initial $\rho$.  Such sets of histories are said to ``decohere''.
We shall defer until Section V a discussion of the precise measure of
the coherence between histories and the implications of decoherence for
time symmetry in quantum mechanics.   We concentrate now on the
theoretical status of the arrow of time exhibited by \eqref{twoone} in the
quantum mechanics of cosmology.

An arrow of time built into a basic quantum mechanics of cosmology may
not (as in the approximate ``Copenhagen'' quantum mechanics of measured
subsystems) be attributed to the thermodynamic arrow of an external
measuring apparatus or larger universe.  In general, these external
objects are not there.
An arrow of
time in the quantum mechanics of cosmology would be a fundamental time
asymmetry in the basic laws of physics.  Indeed, given that,
as we mentioned in the Introduction, the other observed time asymmetries could
all arise from time-symmetric dynamical laws solved with time-asymmetric
boundary conditions, a fundamental arrow of time in the laws
of quantum mechanics could be the only fundamental source of time 
asymmetry in all of physics.

There is no inconsistency between known data and a fundamental
arrow of time in quantum mechanics.  General time asymmetries {\it are} 
exhibited by our universe and there is no evidence suggesting any violation of
causality.  The observed time asymmetries such as the thermodynamic
arrow of time, the arrow of retarded electromagnetic radiation, the
absence of white holes, etc.~{\it  could} all be seen to follow from a
fundamental quantum-mechanical distinction between the past and future.
That is, they could all be seen to arise from a special initial $\rho$
in a quantum-mechanical framework based on \eqref{twoone}.

But might it not be instructive to generalize quantum mechanics so that
it does not so blatantly distinguish past from future?  One could then
investigate a more general class of quantum cosmologies and identify
those that are compatible with the observed time asymmetries.  Even if it
is highly unlikely that ordinary quantum mechanics needs to be replaced
by such a generalization, the generalization can still provide an
instructive way of viewing the origin of time asymmetry in the universe
and provide a framework for discussing tests of the usual assumptions.
We shall discuss in the next section a quantum mechanics that employs
two boundary conditions, one for the past and one for the future, to
give a time-neutral formulation.  Each condition is represented by a
density matrix and the usual theory is recovered when the future density
matrix is proportional to the unit matrix while the one for the past is
much more special.

\section{A Time-Neutral Formulation of Quantum Mechanics for Cosmology}

Nearly thirty years ago, Aharonov, Bergmann, and Lebovitz \cite{ABL64}
showed how to cast the quantum mechanics of measured subsystems into
time-neutral form by considering final conditions as well as initial
ones.  The same type of framework for the quantum mechanics of closed
systems has been discussed by Griffiths \cite{Gri84} and ourselves \cite{Har91}.
In this formulation the probabilities for the individual members of a
set of alternative histories is given by
\begin{subequations}
\label{threeone}
\begin{equation}
p(\alpha_n, \cdots, \alpha_1)= NTr\Bigl[\rho_f P^n_{\alpha_n}(t_n)\cdots
P^1_{\alpha_1}(t_1)\ \rho_i\ P^1_{\alpha_1}(t_1)\cdots
P^n_{\alpha_n}(t_n)\Bigr]\, ,
\label{threeone a}
\end{equation}
where
\begin{equation}
N^{-1}=Tr\left(\rho_f\rho_i\right)\, .
\label{threeone b}
\end{equation}
\end{subequations}
Here, $\rho_i$ and $\rho_f$ are Hermitian, positive operators that we may
conventionally call Heisenberg operators representing
 initial and final conditions for the
universe respectively.  They need not be normalized as density matrices with
$Tr(\rho)=1$ because \eqref{threeone} is invariant under changes of normalization.

The expression \eqref{threeone} for the probabilities of histories is time-neutral.  
There is a density
matrix at both ends of each history.  Initial and final conditions may
be interchanged by making use of the cyclic property
of the trace.  Therefore, the quantum mechanics of closed systems based on \eqref{threeone}
need not have a fundamental arrow of time.  

Different quantum-mechanical theories of cosmology are specified by different
choices for the initial and final conditions $\rho_i$ and $\rho_f$.  For
those cases with $\rho_f\propto I$, where $I$ is the unit matrix, this formulation 
reduces to that
discussed in the previous Section because then \eqref{threeone} coincides with
\eqref{twoone}.  

Of course, the condition for decoherence must also be extended to incorporate initial
and final conditions.  That extension, however, is
straightforward \cite{Gri84, Har91} and will be reviewed briefly in Section V.  The
result is a generalized quantum mechanics in the sense of Refs.~\cite{Har91} and
\cite{GH90b}.

Lost in this generalization is a built-in notion of causality in quantum mechanics. 
Lost also, when  $\rho_f$ is not proportional to $I$, is any notion of a unitarily evolving
``state of the system at a moment of time''.   There is generally no
effective density matrix like $\rho_{\rm eff} (t)$ in (2.4) from which
{\it alone} probabilities for either the future or past could be
computed.  What is gained is a quantum mechanics without a fundamental arrow of
time in which all time asymmetries may arise in particular cosmologies because
of differences between $\rho_i$ and $\rho_f$ or at particular epochs
from their being near to the beginning or the end.
That generalized quantum mechanics
embraces a richer variety of possible universes, allowing for the
possibility of violations of causality and advanced as well as 
retarded effects.  These, therefore, become testable features of the
universe rather than axioms of the fundamental quantum framework.

From the perspective of this generalized quantum mechanics the task of
quantum cosmology is to find a theory of {\it both}  the initial and
final conditions that is theoretically compelling and fits our existing
data as well as possible. Certainly a final condition of indifference,
$\rho_f\propto I$, and a special initial condition, $\rho_i$, 
 seem to fit our
data well, and there is no known reason for modifying them.
  But how accurately is $\rho_f\propto I$
mandated by the data?  What would be the observable consequences of a
completely time-symmetric boundary condition  that is, 
in a sense, the opposite extreme?

Our ability to detect the presence of a final condition differing from
$\rho_f \propto I$ depends on our experimental access to systems whose
behavior today predicted with $\rho_f \not\propto I$ would be measurably
different from the predictions of that behavior with $\rho_f \propto I$.
Loosely speaking, it depends on our finding physical systems which can
``see'' the final condition of the universe today.  In the following we
examine several candidates for such systems, beginning with simple
classical analyses in Section IV and proceeding to more quantum-mechanical
ones in Section V.

\section{Classical Two-Time Boundary Problems}

\subsection{A Simple Statistical Model}

The simplest explanation of the observed
thermodynamic arrow of time is  the asymmetry between a special,
low-entropy,\footnote{For quantitative estimates of how low the initial entropy is, see
\cite{Pen79}.}
 initial condition and a
 maximal-entropy final condition describable as
indifference with respect to final
state (or as no condition at all!). 
Studying deviations of the entropy increase
predicted by statistical mechanics with these boundary conditions from
that predicted with time-symmetric boundary conditions is a
natural way to try to discriminate between the two.  Such
studies were carried out in classical statistical models by by Cocke \cite{Coc67},
Schulman \cite{Sch77}, Wheeler \cite{Whe79}, and others in the late '60s
and early '70s. Schulman, in particular, has written extensively on
these problems both in classical and quantum mechanics \cite{Schff}. We briefly
review such statistical models here. 

Relaxation to equilibrium is a time-symmetric process in a universe
with an underlying dynamics that is time reversal invariant.  Without
boundary conditions, a  
system out of equilibrium is just as likely to have evolved from a state
of higher entropy as it is to evolve to a state of higher entropy.  The
characteristic relaxation time for a system to evolve to equilibrium
depends on the size of the system and the strength of the interactions
that equilibrate it.  Other factors being equal, the larger the system 
the longer the
relaxation time.

There is no simpler instructive model to illustrate the approach
to equilibrium than the Ehrenfest urn model \cite{Kac59}.  For this
reason, it and related models have been much studied in connection with
statistical two-time boundary problems \cite{Coc67, Sch77}.  The model consists
of two boxes and a numbered set of $n$ balls that are distributed between
them.  The system evolves in time according to the following dynamical
rule:  At each time step a random number between 1 and $n$ is produced,
and the ball with that number is moved from the box containing it 
to the other box.
This dynamical rule is time-symmetric.

The fine-grained description of
this system specifies which ball is in which box (a ``microstate'').  An
interesting coarse-grained description involves following just the total
number of balls in each box ( a ``macrostate'') irrespective of {\it which}
balls are in which box.  Let us use this coarse graining to 
 consider an initial condition in which
all the balls are in one box and follow the approach to equilibrium as a
function of the number of time steps, with no further conditions.  Figure
1 shows a numerical calculation of how the entropy averaged over many
realizations of this evolution grows with time to approach its maximum,
equilibrium value.  The relaxation time, obtained either analytically or
numerically, is approximately the total number of balls, $t_{\rm relax}\sim
n$.  If there are no further constraints, the system tends to relax to
equilibrium and remain there.

\begin{figure}[t!]
\begin{widetext}
\parbox[t]{3.5in}{\epsfig{file=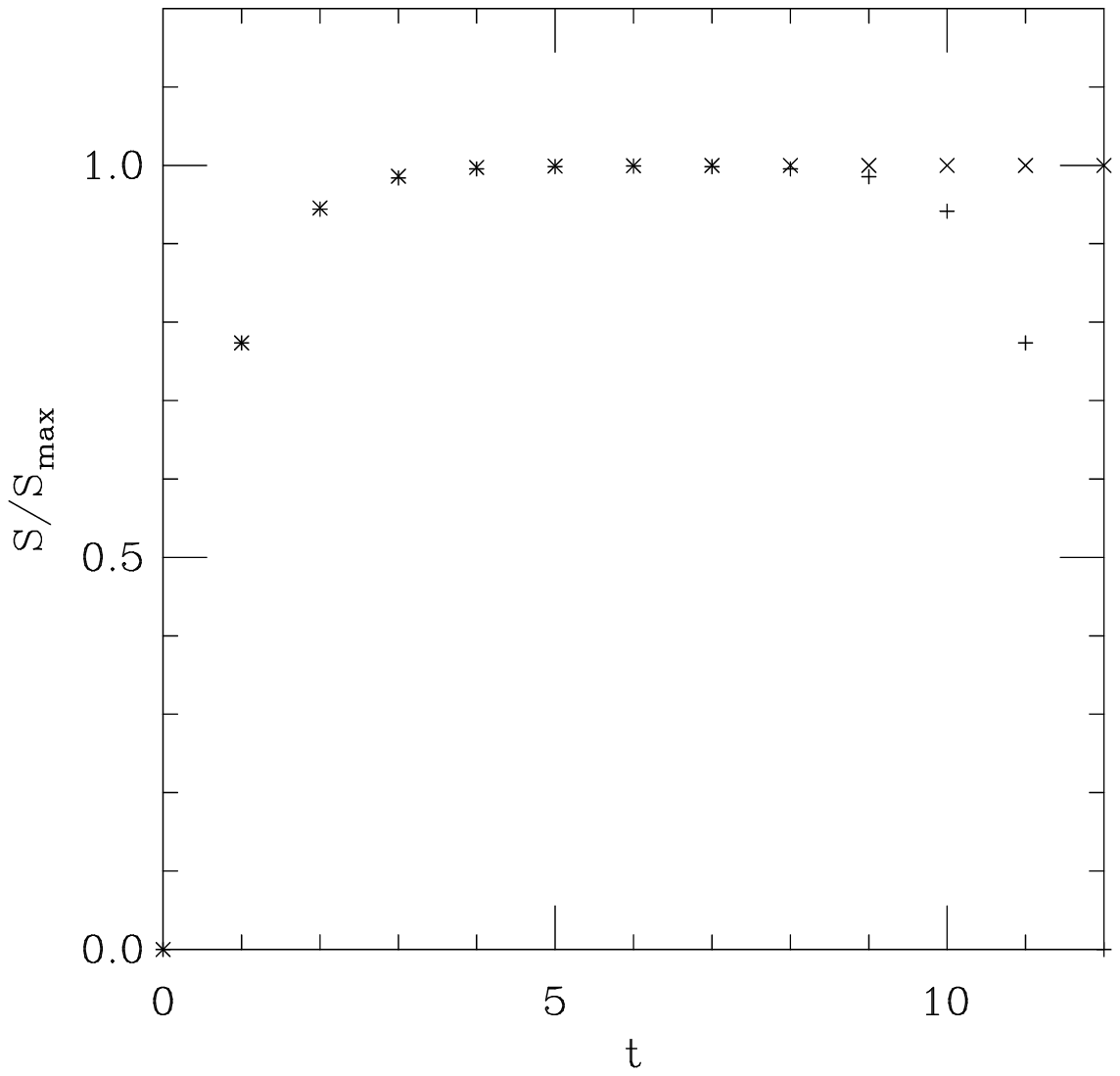, bbllx=39, bblly=39, bburx=586, bbury=558,
clip=true, height=3.5in}}\parbox[t]{3.5in}{
\epsfig{file=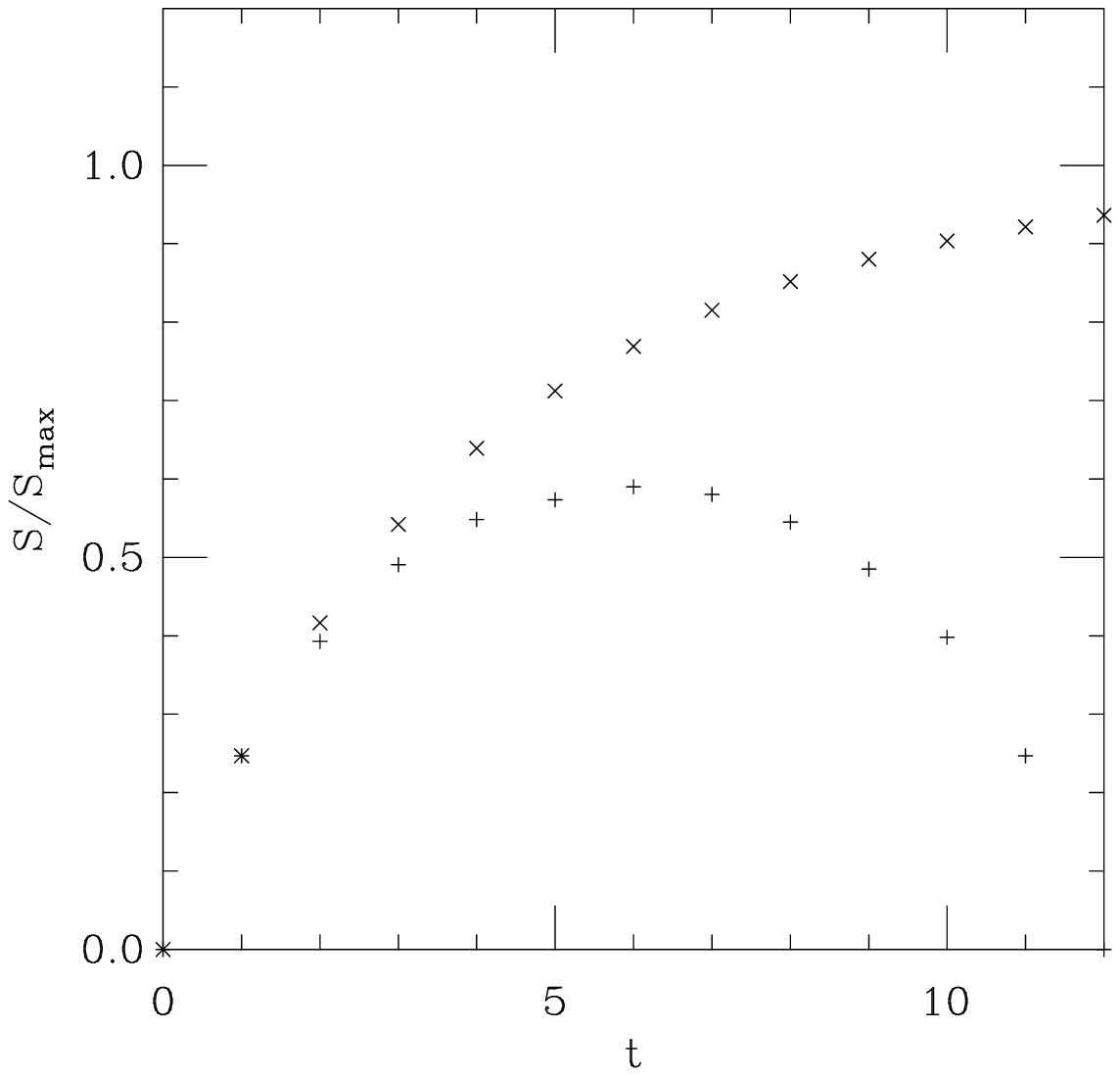, bbllx=39, bblly=39, bburx=586, bbury=558,
clip=true, height=3.5in}}
\small
\setlength{\baselineskip}{.15in}
{Figure 1: Simulation of the approach to equilibrium in
the Ehrenfest urn model.  The entropy of a coarse-grained state in which
only the total number of balls in each box  is followed is the logarithm
number of different ways of distributing the balls between the boxes
consistent with a given total number in each.  This figure shows this
entropy averaged over a large number of different evolutions of the
system for several situations.  These simulations
were carried out by the authors but are no different in spirit from
those discussed by Cocke \cite{Coc67}.\\
The left figure shows the evolution of a 
system of four balls.
  In each case the system starts with all
balls in one box --- a configuration of zero entropy as far from
equilibrium as it is possible to get.  The $\times$'s show  how the
average entropy of 12,556 cases approaches 
equilibrium when there are no further constraints.
  The entropy approaches its equilibrium value in a relaxation
time given approximately by its size, $t_{\rm relax} \sim 4$, and
remains there.  The curve of $+$'s shows the evolution when a time
symmetric final condition is imposed at $T=12$ that all balls have
returned to the one box from whence they started at $t=0$.  A total of 
100,000
evolutions were tried. \\
The average entropy of the 12,556 
cases that met
the final condition is shown.  It is time symmetric about the midpoint,
$T/2=6$.  The initial approach to equilibrium is virtually
indistinguishable from the approach without a final condition because
$t_{\rm relax} \sim 4$ is significantly less than the time $T=12$ at
which the final condition is imposed.  Only within one relaxation time
of the final condition is the deviation of the evolution from the
unconstrained case apparent.\\
The right figure shows the same two cases 
for a larger system of twenty
balls.  The unconstrained approach to equilibrium shown by
the $\times$'s was calculated from the
average of 1010 evolutions and exhibits a relaxation time, $t_{\rm relax} \sim
20$.  The average entropy when time-symmetric boundary conditions are
imposed at $T=12$ is shown in the curve of $+$'s.
10,000,000 evolutions
were tried of which 1010 met the time-symmetric final condition.  (This
demonstrates vividly that it is very  improbable for the entropy of even
a modest size system to fluctuate far from equilibrium as measured by
the $CPU$ time needed to find such fluctuations.)\\ 
The deviation from the
 unconstrained approach to
equilibrium caused by the imposition of a time-symmetric final condition
is significant from an early time of about $t=3$ as the
differences between the $+$'s and the $\times$'s show.
  These models suggest that to detect 
the effects of a 
time symmetric
final condition for the universe we must have access to systems for
which the  
relaxation time is at least comparable to the time difference between
initial and final conditions.}
\end{widetext}
\end{figure}

Consider evolution in the Ehrenfest urn model when a final
condition identical to the initial one is imposed at a later time $T$.
Specifically, construct an ensemble of evolutions consistent with
these boundary conditions by evolving forward from an initial condition
where all the balls are in one box but accepting only those evolutions
where all the balls are back in this box at time $T$.  Figure 1 shows
the results of two such calculations, one for a system with a small number of
balls (where the relaxation time is
significantly smaller than $T$) and the other for a system with 
a larger number of balls
(where it is significantly larger than $T$.)

For both systems the time-symmetric boundary conditions imply a behavior
of the average entropy that is time-symmetric about the midpoint, $T/2$.
For the system with a relaxation time short compared to the time at
which the final condition is imposed, the initial approach to
equilibrium is nearly indistinguishable from that in the case with no 
final condition.  That is because, in equilibrium, the system's
coarse-grained dynamics is essentially independent of its initial {\it
or} final condition.  It, in effect, ``forgets'' both from whence it
started and whither it is going.

By contrast, if the relaxation time is comparable to or greater than the
time interval between initial and final condition, then there will be
significant deviations from the unconstrained approach to equilibrium.
Such systems typically do not reach equilibrium before the effect of the
final condition forces their entropy to decrease.

The evolution of the entropy in the presence of time-symmetric initial
and final conditions must itself be time-symmetric when averaged over many
evolutions, as the simulations in Figure 1 show.  However, in a
statistical theory with time-symmetric boundary conditions the
individual histories need not be time-symmetric.  Figure 2 shows an
example of a single history from an urn model calculation for which the
average
behavior of the entropy is shown in Figure 1.  The ensemble of
histories is time-symmetric by construction; the individual
histories need not be. Since, by definition, we experience 
only one history of the universe, this leaves open the possibility that
the time-asymmetries that we see
could be the result of a statistical fluctuation
in a universe with time-symmetric boundary conditions. In quantum
cosmology, we would not count such a theory of the initial and final
conditions as successful if the fluctuations required were very
improbable. However, in some
examples the magnitude of the fluctuation need not be so very 
large. For instance, consider a classical statistical theory in which
the boundary conditions allow an ensemble of classical histories each
one of which displays an arrow of time and also, with equal probability,
the time-reversed of that history displaying the opposite arrow of time.
The boundary conditions and the 
resulting ensemble are time-symmetric, but predict an observed
time-asymmetry with probability one. Of course, such a theory might
be indistinguishable from one that posited boundary conditions 
allowing just one arrow of time. However, other theoretical
considerations may make it reasonable to consider such proposals, for
example, 
the ``no boundary'' initial condition which is believed to
have this character \cite{Pag85}.  In the subsequent discussion of
time-symmetric cosmological boundary conditions we shall assume that
they predict with high probability {\it some} observable differences from the
usual special initial conditions and final condition of indifference 
and investigate what these are.

\subsection{Classical Dynamical Systems with Two-Time Statistical Boundary
Conditions}

The analysis of such simple classical statistical models with two-time
boundary conditions suggests the behavior of a classical cosmology with
time-symmetric initial and final conditions.  A classical dynamical
system is described by a phase space and a Hamiltonian, $H$.  We write
$z=(q,p)$ for a set of canonical phase-space co\"ordinates.  The
histories of the classical dynamical system are the phase-space curves
$z(t)$ that satisfy Hamilton's equations of motion.

A statistical classical dynamical system is described by a distribution
function on phase space $\rho^{cl}(z)$  that gives the probability of
finding it in a phase-space volume.  For analogy with quantum mechanics
it is simplest to use a ``Heisenberg picture'' description in which the
distribution function does not depend explicitly on time but
time-dependent co\"ordinates on phase space are used to describe the
dynamics.  That is, if $z_0$ is a set of canonical co\"ordinates at time
$t=0$, a set $z_t$ appropriate to time $t$ may be defined by
$z_t=z_0(t)$ where $z_0(t)$ is the classical evolution of $z_0$
generated in a time $t$ by the Hamiltonian $H$.  The statistical system
at time $t$ is then distributed according to the function $\rho^{cl}$
expressed in terms of the co\"ordinates $z_t$, viz. $\rho^{cl}(z_t)$. The
distributions $\rho^{cl}(z_t)$ and $\rho^{cl}(z_{t^\prime})$ will
therefore typically have {\it different} functional forms.  

An ensemble of histories distributed according to the probabilities of a
statistical classical dynamical system with boundary conditions at two
times $t_i$ and $t_f$ might be constructed as follows:  Evolve a large
number of systems distributed according to the initial distribution
function $\rho^{cl}_i(z_i)$ forward from time $t_i$ to time $t_f$.  If a
particular system arrives at time $t_f$ in the phase space volume
$\Delta v$ centered about point $z_f$, select it for inclusion in the
ensemble with probability $\rho^{cl}_f (z_f)\Delta v$ where  $\rho^{cl}_f$
is the distribution function representing the final boundary
condition.  Thus, if $\rho^{cl}_i$ and $\rho^{cl}_f$ are referred to a common
set of phase-space co\"ordinates, say $z_t$, the time-symmetric ensemble
of systems will be
distributed according to the function
\begin{subequations}
\label{fourone}
\begin{equation}
\bar\rho^{cl}(z_t) = N\rho^{cl}_f(z_t)\rho^{cl}_i(z_t)\, ,
\label{fourone a}
\end{equation}
where
\begin{equation}
N^{-1} = \int dz_t\ \rho^{cl}_f (z_t) \rho^{cl}_i(z_t)\, .
\label{fourone b}
\end{equation}
\end{subequations}
Referred to the initial time, \eqref{fourone} has a simple interpretation:  Since
classical evolution is unique and deterministic the selection at the
final time could equally well be carried out at the initial time with
$\rho^{cl}_f$ evolved back to the initial time.  The distribution
 $\bar\rho^{cl}$ is the
result.

We now discuss the relation between $\rho^{cl}_i$ and $\rho^{cl}_f$ that
is necessary for the probabilities of this classical cosmology to be
symmetric about a moment of time.  Take this time to be $t=0$ and
introduce the operation, ${\cal T}$, of time-reversal about it.
\begin{equation}
{\cal T} \rho^{cl} (q_0, p_0) \equiv \rho^{cl} (q_0, -p_0)\, .
\label{fourtwo}
\end{equation}
If we assume that the Hamiltonian is itself time-reversal invariant
\begin{equation}
H(q_0, p_0, t) = H (q_0,-p_0, -t)\, ,
\label{fourthree}
\end{equation}
this implies
\begin{equation}
{\cal T}\rho^{cl}(q_t,p_t) = \rho^{cl}(q_{-t},-p_{-t})\, .
\label{fourfour}
\end{equation}
The distribution function \eqref{fourone} may then be conveniently rewritten
\begin{equation}
\bar\rho^{cl} (q_t, p_t) = N\rho^{cl}_i (q_t, p_t) {\cal T} \rho^{cl}_f
(q_{-t}, -p_{-t})\, .
\label{fourfive}
\end{equation}
A relation between $\rho^{cl}_i$ and $\rho^{cl}_f$ sufficient to imply
the time-symmetry of the distribution $\bar \rho^{cl}$ is now evident,
namely
\begin{equation}
\rho^{cl}_f (q_t, p_t) = {\cal T}^{-1}
\rho^{cl}_i (q_t, p_t)\, .
\label{foursix}
\end{equation}
The final condition is just the time-reversed version of the initial one.

The imposition of time-symmetric statistical boundary
conditions on a classical cosmology means in particular that the entropy
must behave time-symmetrically if it is computed utilizing a coarse graining
that is itself time-symmetric.  The entropy of the final distribution
must be the same as the initial one.  The thermodynamic arrow of time
will run backwards on one side of the moment of time symmetry as
compared to the other side.  This does not mean, of course, that the
histories of the ensemble need be individually time-symmetric, as the
example in Figure 2 shows.  In particular, subsystems with
relaxation times long compared to the interval between initial and final
conditions might have non-negligible probabilities for fluctuations from
exactly time-symmetric behavior.  There would appear to be no
principle, for example, forbidding us to live on into the recontracting 
phase of the
universe and see it as recontracting.  It is just that as time
progressed events judged to
be unexpected on the basis of just an initial condition would happen
with increasing frequency.  It is by the frequency of such unexpected
events that we could detect the existence of a final condition.  

Could we infer the existence of a time-symmetric final condition
for the universe from the deviations that it would imply for the
approach to equilibrium that would be expected were there no such final
condition?  The statistical models reviewed in  above suggest that to
do so we would need to study systems with relaxation times comparable 
to or longer than the
lifetime of the universe between the ``big bang'' and the ``big crunch''.  If
the lifetime of the universe is comparable to the present age of the
universe from the ``big bang'', then we certainly know such systems.
Systems of stars such as galaxies and clusters provide ready examples.
Any single star, with the ambient radiation, provides another as long
as the star's temperature is above that of the cosmic background radiation.
Black holes with lifetime to decay by the Hawking radiation longer
than the Hubble time are further examples.  Indeed, from the point of
view of global cosmological geometry, the singularities contained within
black holes can be considered to be parts of the final singularity of
the universe, where a
final condition would naturally be imposed \cite{Pen78}.  The singularities of 
detectable black holes
may be the parts of this final singularity  closest to us. On smaller scales,
samples of radioactive material with very long half-lives may be other
such examples and Wheeler \cite{Whe79} has discussed experiments utilizing
them to search for a time-symmetric final
condition.
We may hope, as mentioned above, that the evolving collective complex adaptive
system of which we are a part could be such a long-lasting phenomenon!

\begin{figure}[t]
\begin{widetext}
\centerline{\parbox[t]{3.5in}{\epsfig{file=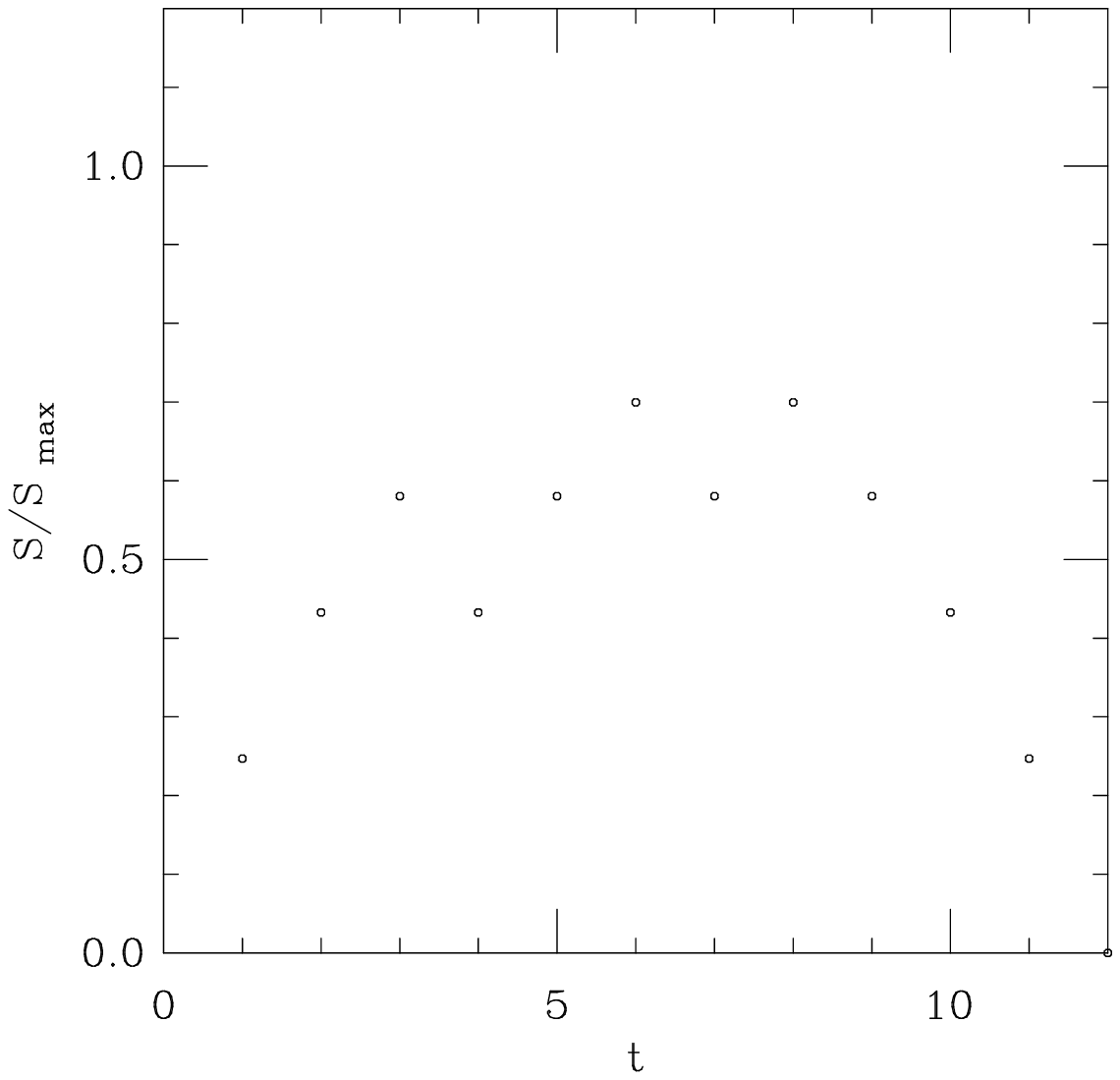, bbllx=39, bblly=39, bburx=586, 
bbury=558, clip=true, height=3.5in}}}
\small
\setlength{\baselineskip}{.15in}
{Fig 2. An individual evolution in the urn model with
time-symmetric initial and final conditions.  The history of the entropy
averaged over many evolutions must clearly be time-symmetric in a
statistical theory with time-symmetric initial and final conditions as
Figure 1 shows.  However, the individual evolutions need not be
separately time-symmetric.  This figure shows the entropy for the case
of twenty balls in the first evolution among the 10,000,000 described
in Figure 1 that met the time-symmetric final condition.  It is not
time-symmetric.  For systems such as this with relaxation time, $t_{\rm
relax}$, larger than the time between initial and final conditions,
significant deviations from exact time symmetry may be expected.}
\end{widetext}
\end{figure}

However, if the lifetime of the universe is much {\it longer} than the
present age from the ``big bang'', then it might be much more difficult to
find systems that remain out of equilibrium long enough for their
initial approach to equilibrium to be significantly affected by a
time-symmetric final condition.  That could be the case with the
$\Omega$-near-one universe that would result from a rapid initial inflation.  If
its lifetime were long enough, we might never be able to detect the
existence of a time-symmetric final condition for the 
universe.

The lifetime of our classical universe obeying the Einstein equation is,
of course, in principle determinable from present observations (for
example, of the Hubble constant, mean mass density, and deceleration
parameter).  Unfortunately we do not have enough data to distinguish
observationally 
between a lifetime of about twenty-five billion years and an infinite
lifetime. Very long lifetimes are not only consistent with observations,
but also, as we now describe, are suggested theoretically by quantum
cosmology as a consequence of inflation.

The quasiclassical cosmological evolution that we observe should
be, on a more fundamental level,
 a likely prediction of a quantum theory of the
universe and its boundary conditions.  We shall discuss time symmetry in
the context of quantum cosmology in later sections, but for the present
discussion we note that, 
in quantum cosmology, the probabilities for different lifetimes of the
universe are predictable from a theory of its initial and final
conditions.  That is because in a quantum theory that includes
gravitation the geometry
of spacetime, including such features as the time between the  ``big bang'' and
a ``big crunch'', if any, is quantum-probabilistic.  

A quantum state that predicts quasiclassical behavior does not typically
predict a unique classical history but rather an ensemble of possible
classical histories with different probabilities.  This is familiar from
wave functions of WKB form, which do not predict single classical
trajectories but only the classical connection between momentum and
classical action.
Similarly, in the quantum mechanics of closed cosmologies,  we expect
a theory of quantum boundary conditions to determine an ensemble of
different classical cosmological geometries with different
probabilities.\footnote{For further discussion see {\it e.g.}~\cite{Har91}.}  The
geometries in the ensemble will have {\it different times} between the
``big bang'' and the ``big crunch'' because in quantum gravity that time is a
dynamical variable and not a matter of our choice.  In this way, the
probability distribution of lifetimes of the universe becomes
predictable in quantum cosmology.

Cosmological theories that predict inflation lead to very large expected
values for the lifetime of the universe; and inflation seems to be
implied by some currently interesting theories of the
boundary conditions of the universe. The question has been analyzed only
for theories with a special initial condition, such as the ``no-boundary
proposal'' and the ``tunneling-from-nothing proposal''.
Analyses by Hawking and Page \cite{HP88}, Grishchuk and Rozhansky \cite{GR90}, and
Barvinsky and Kamenshchik \cite{BK90} suggest very large expected lifetimes
for the ``no-boundary
proposal''.  Analyses by Vilenkin \cite{Vil88} and by Miji\'c, Morris, and
Suen \cite{MMS89} do likewise for the ``tunneling-from-nothing'' case.

\subsection{Electromagnetic Radiation}

The above discussion suggests that in order to probe the nature of a
non-trivial final condition, one should study processes today that are
sensitive to that final condition no matter how far in the future it is
imposed.  At the conference, P.C.W.~Davies suggested that
electromagnetic radiation might provide such a mechanism for ``seeing''
a final condition in the arbitrarily far future in realistic
cosmologies.  In an approximately static and homogeneous cosmology,
radiation must travel through ever more material the longer the time
separation between initial and final conditions.  For sufficiently large
separations, the universe becomes opaque to the electromagnetic
radiation necessary to  probe the details of the
final condition directly.  However, in an expanding
universe the dilution of matter caused by the expansion competes with
the longer path length as the separation between initial big bang and
final big crunch becomes longer and longer.  Davies and Twamley \cite{DT93}
show that, under reasonable conditions, the expansion wins and that the
future light cone is transparent to photons all the way to a distance
from the final singularity comparable to ours from the big bang.  

Transparency of the forward light cone raises the possibility of
constraining the final condition by present observations of
electromagnetic radiation and perhaps ruling out time-symmetric
boundary conditions.  Partridge \cite{Par73} has actually carried out
experiments which could be interpreted in this way and Davies and
Twamley discuss others. The following is an example of a further
argument of a very direct kind.

Suppose the universe to have initial and final classical distributions
that are time-symmetric in the sense of \eqref{foursix}. Suppose further that these
boundary conditions imply with high probability an initial epoch with
stars in galaxies distributed approximately homogeneously and a similar
final epoch of stars in galaxies at the symmetric
time.  Consider the radiation emitted from a particular star in
the present epoch.  If the universe is transparent, it is likely to
reach the final epoch without being absorbed or scattered.  There it may
either be absorbed in the stars or proceed past them towards the
final singularity.  If  a significant fraction of the radiation proceeds
past, then by time-symmetry we should expect a corresponding amount of
radiation to have been emitted from the big bang.  Observations of the
brightness of the night sky could therefore constrain the possibility
of a final boundary condition time-symmetrically related to the initial
one. The alternative
that the radiation is completely absorbed in future stars implies
constraints on present emission that are probably inconsistent with
observation because the total cross section of future stars is only a
small fraction of the whole sky, as it is today.\footnote{Thanks are
due to D.~Craig for discussions of this example.}

By such arguments, made quantitative, and extended to neutrinos,
gravitational and other forms of radiation, we may hope to constrain the
final condition of the universe no matter how long the separation
between the big bang and the big crunch.

\section{Hypothetical Quantum Cosmologies with Time Symmetries}

\subsection{$CPT$- and $T$-Symmetric Boundary Conditions}

The time-neutral generalized quantum mechanics with initial and final
conditions developed in Section III permits the construction of model
quantum cosmologies that exhibit symmetries with respect to reflection
about a moment of time.  By this we mean that the probabilities given by
\eqref{threeone} for a set of alternative histories are identical to those of the
symmetrically related set.  This section explores the relations between
$\rho_f$ and $\rho_i$ and the conditions on the Hamiltonian under which
such symmetries exist.

$CPT$-symmetric universes are the most straightforward to implement
because local field theory in flat spacetime is invariant under $CPT$.  We
expect $CPT$ invariance as well for field theories in curved cosmological
spacetimes such as closed Friedmann universes that are symmetric
under a space inversion and symmetric about a moment of time.

To construct a $CPT$-invariant quantum cosmology, choose the origin of time
so that the time reflection symmetry is about $t=0$.  Let $\Theta$
denote the antiunitary $CPT$ transformation and for simplicity consider
alternatives $\{P^k_{\alpha_k} (t_k)\}$ such that their
$CPT$ transforms,  $\{\widetilde P^k_{\alpha_k}(-t_k)\}$, are given by
\begin{equation}
\widetilde P^k_{\alpha_k}(-t_k) = \Theta^{-1} P^k_{\alpha_k} (t_k)
 \Theta\, .
\label{fiveone}
\end{equation}
A $CPT$-symmetric universe would be one in which the probabilities of
histories of alternatives at times $t_1<t_2< \cdots <t_n$ would be
identical to the probabilities of the $CPT$-transformed histories of
alternatives at times $-t_n<\cdots <-t_2<-t_1$. Denote by 
$C_\alpha$ the
string of projection operators representing one history
\begin{equation}
C_\alpha = P^n_{\alpha_n} (t_n) \cdots P^1_{\alpha_1}(t_1)\, ,
\label{fivetwo}
\end{equation}
and by $\widetilde C_\alpha$ 
the corresponding string of $CPT$-transformed alternatives written in 
standard time order with the earliest alternatives to the right
\begin{subequations}
\label{fivethree}
\begin{equation}
\widetilde C_\alpha \equiv 
\widetilde P^1_{\alpha_1}(-t_1)\cdots \widetilde 
P^n_{\alpha_n} (-t_n)\, .
\label{fivethree a}
\end{equation}
Thus, 
\begin{equation}
\widetilde C_\alpha = \Theta^{-1}C^{\dagger}_\alpha \Theta \,  .
\label{fivethree b}
\end{equation}
\end{subequations}
The requirement of $CPT$ symmetry is then, from \eqref{threeone},
\begin{equation}
Tr \left(\rho_fC_\alpha \rho_i C^\dagger_\alpha\right) = 
Tr\left(\rho_f\widetilde C_\alpha
\rho_i \widetilde C^\dagger_\alpha\right)\, .
\label{fivefour}
\end{equation}
Using \eqref{fiveone}, \eqref{fivethree}, the cyclic property of the trace, 
and the identity
$Tr[A \Theta^{-1} B \Theta]=Tr[B^\dagger \Theta A^\dagger \Theta^{-1}]$
following from the antiunitarity of $\Theta$, the right hand side
of \eqref{fivefour} may be rewritten to yield the following form of the requirement
of $CPT$ symmetry
\begin{equation}
Tr\left(\rho_f C_\alpha\rho_i C^\dagger_\alpha\right) = Tr
\left(\Theta \rho_i \Theta^{-1} C_\alpha \Theta \rho_f \Theta^{-1}
C^\dagger_\alpha\right)\, .
\label{fivefive}
\end{equation}
Evidently a sufficient condition for a $CPT$-symmetric universe is that
the initial and final conditions be $CPT$ transforms of each other:
\begin{equation}
\rho_f = \Theta\rho_i \Theta^{-1}
\label{fivesix}
\end{equation}
because acting on Bose operators $\Theta^2$ is effectively unity
 and  as a consequence of (5.6) $\rho_i = \Theta \rho_f
\Theta^{-1}$ .

As stressed by Page [34], a $CPT$-symmetric universe can also be realized
with within the usual formulation of quantum mechanics with an 
initial $\rho_i$ and a final $\rho_f=I$, provided the 
$\rho_i$ representing the condition at the initial instant is 
$CPT$-{\it invariant} about some time in the future. Thus, initial and
final conditions that are not related by (5.6) do not {\it necessarily}
imply differing probabilities for sets of histories connected by $CPT$.
Further, as discussed in the previous section, both ways of realizing
a $CPT$-symmetric universe can, with appropiate kinds of initial and final
conditions and coarse-graining, lead to sets of histories in which each
individual member is $CPT$-{\it a}symmetric about the moment of
symmetry. 
Thus, neither are $CPT$-symmetric boundary conditions {\it
necessarily} inconsistent with arrows of time that extend consistently
over the whole of the universe's evolution. 

A universe is time-symmetric about a moment of time if the
probabilities of any set of alternative histories are identical to those
of the time-inverted set.
The relation between initial and final conditions necessary for a purely
time-symmetric universe is analogous to that for a $CPT$-symmetric one
 and derived in the same way.
However, we cannot expect boundary conditions to impose 
time symmetry if the
Hamiltonian itself distinguishes past from future. We must assume 
that the Hamiltonian is symmetric under
time inversion, ${\cal T}$,
\begin{equation}
{\cal T}^{-1} H (t) {\cal T} = H(-t)\, .
\label{fiveseven}
\end{equation}
Given \eqref{fiveseven},
a time-symmetric universe will result if the initial and final
conditions are related by time inversion:
\begin{equation}
\rho_f={\cal T} \rho_i {\cal T}^{-1}\, .
\label{fiveeight}
\end{equation}

For realistic quantum cosmologies, the time-neutral quantum mechanics of
universes in a box, described in Section III, must be generalized to allow
for significant quantum fluctuations in spacetime geometry, and notions
of space and time inversion must be similarly generalized.  A sketch of a
generalized quantum mechanics for spacetime can be found in \cite{Har91} and
discussions of time inversion in the quantum mechanics of cosmology in
\cite{Pag93}, \cite{Pag85}, and \cite{Haw85}.

\subsection{$T$ Violation in the Weak Interactions}

The effective Hamiltonian describing the weak interaction on accessible
energy scales is not $CP$-invariant.  As a consequence of the $CPT$
invariance of field theory it is also not $T$-invariant.  
$T$ violation of this kind is a small effect in laboratory experiments but
is thought to be  of central
importance in the evolution of the matter content of the universe.  It
is believed to be responsible, together with the non-conservation  
of baryons,  for the emergence of a matter-dominated
universe from an initial equality of matter and
antimatter, as originally pointed out by 
Sakharov.\footnote{Ref.~\cite{Sak79}.  For an accessible recent 
review of these ideas see
\cite{KT90}.}  Can the symmetric universes just discussed be consistent with
this effective $T$ violation in the weak interaction?

The violation of time-inversion symmetry that we observe in the
effective weak interaction Hamiltonian could arise in three ways:
First, it could be the result of $T$ violation in the fundamental
Hamiltonian.  Second, it could arise throughout the universe even if the
fundamental Hamiltonian were time-inversion-symmetric, from asymmetries
in the cosmological boundary conditions of the universe.  Third, it
could be an asymmetry of our particular epoch and spatial location arising
dynamically in extended domains from a time-inversion symmetric
Hamiltonian and boundary conditions.  We shall now offer a few comments
on each of these possibilities.

If the fundamental Hamiltonian is time-inversion asymmetric, then we
cannot expect a time-symmetric universe, as we have already discussed.
One could investigate whether such a fundamental time asymmetry is the
source of the other observed time asymmetries.  So far such an approach
has neither been much studied nor shown much promise.

Even though a $T$-symmetric universe is inconsistent with a
$T$-asymmetric fundamental Hamiltonian, a $CPT$-symmetric universe could
be realized if the initial and final density matrices were related by
\eqref{fivesix}.  That is because a field-theoretic Hamiltonian is always
$CPT$-symmetric even if it is not $T$-symmetric. But
$CPT$ symmetry needs to be reconciled with the observed
matter-antimatter asymmetry over large domains\footnote{For a classic
review of the observational evidence that there is a matter-antimatter
asymmetry over a domain at least the size of the local group of galaxies
see \cite{Ste76}.}  of the universe and the  
classical behavior of their matter content.  If the universe is 
{\it homogeneously} matter-dominated 
now, then $CPT$ symmetry would imply that it will be homogeneously 
 antimatter-dominated at the time-inverted epoch in the future.  
What evolution of
the present universe could lead to such an inversion?
One possibility is a universe that lasts much longer than the proton
lifetime.\footnote{We owe this suggestion to W.~Unruh.}

There is no evidence for $CP$ violation in the basic dynamics of
superstring theory.  If it is the correct theory, the effective $CP$
violation in the weak interaction in four dimensions has to arise in the
course of compactification or from some other form of spontaneous
symmetry breaking. From the four-dimensional point of view, which we are
taking for convenience in this article, this would correspond to having
a non-zero expected value of a $CP$-odd quantity.
Then, as discussed above,
it is possible to investigate time-symmetric universes with
initial and final conditions related by \eqref{fiveeight}.  An effective
$CP$ violation could arise from $CP$ asymmetries of the initial or final
states or both.  Typical theories of these boundary conditions relate
them to the Hamiltonian or an equivalent action.  Each density matrix,
$\rho_i$ or $\rho_f$, may either inherit the symmetries of the
fundamental Hamiltonian or be an asymmetrical member of a symmetrical
family of density matrices determined by it.  This is the case, for
example, with ``spontaneous symmetry breaking'' of familiar field theory
where there are degenerate candidates for the ground state not individually
 symmetrical under the
symmetries of the Hamiltonian.  Before discussing the
possibility of effective $CP$ violation in time-symmetric universes, let
us review how an effective $CP$ violation can arise in familiar field
theory and in usual quantum cosmology with just an initial condition.

Effective $CP$ violation can arise in field theory even when the
fundamental Hamiltonian is $CP$-invariant, provided there is a non-vanishing
vacuum expected value of a $CP$-odd field $\phi(\vec x, t)$ \cite{Lee74}, {\it i.e.}~one  
such that
\begin{equation}
\phi(-\vec x, t) = -({\cal CP})^{-1} \phi (\vec x, t) ({\cal CP})\, .
\label{fivenine}
\end{equation}
Usually the vacuum state $|\Psi_0\rangle$ inherits the symmetry of the
Hamiltonian that determines it and the vacuum expected value of a
$CP$-odd field would vanish if the Hamiltonian is $CP$-invariant.
However, if there is a symmetrical family of
degenerate candidates for the ground state that are individually 
not $CP$-invariant, then
the expected value
\begin{equation}
\langle \phi (\vec x, t)\rangle = Tr\left[\phi(\vec x, t) |\Psi_0\rangle
\langle\Psi_0|\right]
\label{fiveten}
\end{equation}
may be  non-zero for the physical vacuum.

Similarly, in usual quantum cosmology with just an initial condition $\rho_i$,
a non-zero value of
\begin{equation}
\langle\phi (\vec x, t)\rangle= Tr\left[\phi (\vec x, t)
\rho_i\right]\label{fiveeleven}
\end{equation}
can lead to effective $CP$ violation.  The ``no-boundary'' wave function
of the universe \cite{HH83} is the generalization of the flat space notion of
ground state, {\it i.e.}~vacuum, to the quantum mechanics of closed 
cosmological spacetimes.
The ``no-boundary'' prescription with matter theories that would lead to
spontaneous $CP$ violation in flat space thereby becomes an interesting
topic for investigation.  In such situations, we expect the
``no-boundary'' construction to yield a $CP$-symmetric set of possible wave
functions for the universe that are individually $CP$-asymmetric.

We now turn to effective $CP$ violation in time-symmetric universes with
initial and final states related by \eqref{fiveeight}.  An expected value for a 
field is defined when probabilities are assignable to its alternative
values --- that is, when there is decoherence among the alternatives.  
The requirements
of decoherence will be discussed in the next Section.  They are
automatically satisfied for alternatives at a single moment of time when
$\rho_f\propto I$ but they are
non-trivial when $\rho_f$ is non-trivial.  We have
not analyzed the circumstances in which the values of the field decohere
but we assume  those circumstances to obtain 
here so that the expectation value of the field may
be defined.

A consequence of decoherence and the probability formula \eqref{threeone} is the
validity of two
equivalent expressions for the expected value of the field that are
analogous to \eqref{fiveten} and \eqref{fiveeleven}:
\begin{equation}
\langle\phi(\vec x, t) \rangle = NTr\left[\rho_f \phi (\vec x,t) \rho_i
\right] = NTr \left[\rho_i \phi (\vec x,t)\rho_f\right]\, .
\label{fivetwelve}
\end{equation}
These are demonstrated in the Appendix.  The symmetry between the
initial and final conditions in \eqref{fivetwelve} can be understood from the fact
that it is not probabilities at one moment of time that distinguish the
future from the past.
We shall now show that for a $CP$-odd field this expected value is
odd under time inversion for a time-symmetric universe.  

We carry over
from flat space field theory the assumption that we are dealing with a
$CPT$-even field $\phi (\vec x, t)$.  In flat space that is necessary if
the field is to have a non-vanishing vacuum expected value.
The $CPT$ invariance of field theory then means that it is possible to choose
a (real) representation of $\phi(\vec x,t)$ such that
\begin{equation}
\phi(-\vec x, -t) = ({\cal CPT})^{-1} \phi(\vec x, t)
 ({\cal CPT})\, .
\label{fivethirteen}
\end{equation}
Therefore, since $\phi(\vec x, t)$ is $CP$-odd it must be $T$-odd and then
\begin{equation}
\langle\phi(\vec x, -t)\rangle= -Tr\left[\rho_f {\cal T}^{-1}\phi (\vec x,t)
{\cal T} \rho_i\right]\, .
\label{fivefourteen}
\end{equation}
But if $\rho_i$ and $\rho_f$ are related by \eqref{fiveeight} this relation may be
written
\begin{eqnarray}
\langle \phi(\vec x, -t)\rangle & = & -Tr\left[{\cal T}^{-1} \rho_i {\cal T}
{\cal T}^{-1} \phi (\vec x, t) {\cal T} {\cal T}^{-1} \rho_f {\cal
T}\right]\nonumber\\
& = & -Tr\left[\rho_i \phi (\vec x, t)\rho_f\right] = -\langle \phi (\vec
x, t)\rangle\, . 
\label{fivefifteen}
\end{eqnarray}
The conclusion is that it is possible to choose initial and final
conditions so that a universe is time-symmetric and has a non-vanishing
expected value of a $CP$-odd field.  That expected value is odd in
time (the correct time-{\it symmetric} behavior for a $T$-odd field.)
As a consequence the sign of $CP$ violation would be opposite on opposite
sides of the moment of time symmetry and the magnitude of $CP$ violation
would decrease on cosmological time scales as the moment of time symmetry
is approached.  The $CP$ violation in the early universe might well be
larger than generally supposed and Sakharov's mechanism 
for the generation of the baryons more
effective.  However, if the moment of time symmetry is far in our
future, then such variation in the strength of $CP$ violation would be
small and it would be difficult to distinguish this time-symmetric
situation from the kind of $CP$ violation that arises from  just an 
initial condition as discussed above.

In the class of time-symmetric universes just discussed, $CP$ violation
arises from initial and final conditions that are not $CP$-symmetric.
However, 
an effective $CP$ violation could also exist in our epoch, in local spatial
domains, even if both Hamiltonian and initial and final states were
$CP$-symmetric:
\begin{equation}
H=({\cal CP})^{-1} H({\cal CP})\, ,
\ \rho_i= ({\cal CP})^{-1} \rho_i ({\cal CP})\, ,
\ \rho_f=({\cal CP})^{-1} \rho_f ({\cal CP})\, .
\label{fivesixteen}
\end{equation}
Dynamical mechanisms would need to exist that make likely the existence
of large spacetime domains in which $CP$ is effectively broken, say by
the expected value of a $CP$-odd field that grows to be homogeneous
over such a domain.  In such a picture the set of histories of the
universe would be overall $CP$-symmetric and $T$-symmetric, as follows
from \eqref{fivesixteen}.
Individual histories would display effective $CP$ violation in domains
with sizes and durations that are quantum-probabilistic.  If very large
sizes and durations were probable it would be difficult to distinguish
this kind of mechanism from any of those discussed above.

Overall matter-antimatter symmetry would be expected for such universes
with matter or anti-matter predominant only in local domains.  Their
size must therefore be larger than the known scales on which matter is
dominant \cite{Ste76}. The 
calculation of the probabilities for these  sizes and durations thus 
becomes an important
question in such pictures.  An extreme example occurs in the proposal
of Linde \cite{Lin86}, in which such domains are far larger than the present
Hubble radius.

\section{The Limitations of Decoherence and Classicality}

As we mentioned in Section II, the quantum mechanics of a closed system
such as the universe as a whole predicts probabilities only for those
sets of alternative histories for which there is negligible interference
between the individual members in the set.  Sets of histories that
exhibit such negligible interference as a consequence of the Hamiltonian
and boundary conditions are said to decohere.  A minimal requirement on
any theory of the boundary conditions for cosmology is that the universe
exhibit a decoherent set of histories that corresponds to the
quasiclassical domain of everyday experience.  This requirement places
significant restrictions on the relation between $\rho_i$ and $\rho_f$
in the generalized quantum mechanics for cosmology, as we shall now
show. 

\subsection{Decoherence}

Coherence between individual histories in an exhaustive 
set of coarse-grained histories is measured by the decoherence
functional [20].  This is a complex-valued functional on each pair of
histories in the set. If the cosmos is replaced by a box, so that
possible complications from quantum gravity disappear, then   
 individual coarse-grained histories are
specified by
sequences of alternatives $\alpha =
(\alpha_1, \cdots, \alpha_n)$ at discrete moments of time,
$t_1,\cdots,t_n$.  The decoherence functional for the case of two-time
boundary conditions is given by \cite{Gol62}
\begin{equation}
D(\alpha^\prime, \alpha) = NTr \left[\rho_f C_{\alpha^\prime}
\rho_iC^\dagger_\alpha\right]\, .
\label{sixone}
\end{equation}
A set of histories decoheres when the real parts of the
 ``off-diagonal'' elements of
the decoherence functional --- those between two histories with 
any $\alpha_k \not= \alpha^\prime_k$ --- vanish to sufficient accuracy.
As first shown by Griffiths \cite{Gri84}, this is the necessary and sufficient
condition that the probabilities \eqref{threeone}, which are the ``diagonal''
elements of $D$, satisfy the sum rules defining probability theory.

The possibility of decoherence is limited by the choice of initial and 
final density
matrices $\rho_i$ and $\rho_f$.  To see an example of this, consider the
case in which both are pure,
$\rho_i = |\Psi_i><\Psi_i|$ and $\rho_f = |\Psi_f><\Psi_f|$.
The decoherence functional would then factor:
\begin{equation}
D(\alpha^\prime,\alpha) = N <\Psi_f|C_{\alpha^\prime} |
\Psi_i><\Psi_i|C_\alpha | \Psi_f>\, ,
\label{sixtwo}
\end{equation}
where $N$ now is $|<\Psi_i|\Psi_f>|^{-2}$.
In this circumstance the requirement that the real part of $D$ vanish
for $\alpha^\prime \not= \alpha$ can be satisfied only if there are at most
two non-vanishing quantities 
$<\Psi_i|C_\alpha |
 \Psi_f>$, with 
phases differing
by $90 {^\circ}$, giving at
most two histories with non-vanishing probabilities!  Thus 
initial and final states that are both pure, such as those corresponding 
to a ``wave
function of the universe'', leads to a highly unorthodox quantum
mechanics in which there are only one or two coarse-grained histories.
All the apparent accidents of quantum mechanics would be 
determined\footnote{This situation is closely related to the one
described by L.~Schulman \cite{Sch86}.} by
the boundary conditions $\rho_i$ and $\rho_f$.  The usual idea of a
simple $\rho_i$ (or $\rho_i$ and $\rho_f$), with the algorithmic complexity
of the universe contained almost entirely in the throws of the quantum
dice, would here be replaced by a picture in which the algorithmic
complexity is transferred to the state vectors $|\Psi_i\rangle$ and 
$|\Psi_f\rangle$.
Presumably these would be described by a simple set of rules plus a huge
amount of specific information, unknowable except by experiment and
described in practice by a huge set of parameters with random values. 

This bizarre situation refers to the use of a pure $\rho_i$ and a pure
$\rho_f$, whether or not there is any kind of time symmetry relating
them.

\subsection{Impossibility of a Universe with $\rho_f = \rho_i$.}

We shall now give a very special example of a relation between $\rho_i$
and $\rho_f$, stronger than time symmetry,  that is inconsistent 
with the existence of a quasiclassical
domain.  More precisely we shall show that in the extreme case 
\begin{equation}
\rho_f = \rho_i \equiv \rho
\label{sixthree}
\end{equation}
only sets of histories exhibiting trivial
dynamics can exactly decohere.  This condition means that $\rho_f$
has the same form when expressed in terms of the initial fields
$\phi(\vec x, t_0)$ as $\rho_i$ does. Such a situation could arise if,
in addition to time symmetry, we had $\rho_i$ and $\rho_f$ separately,
individually time-symmetric and with  
effectively no time difference
between the initial and final conditions. 
We know of no theoretical reason to
expect such a situation, but it does supply an example that leads to a
contradiction with experience.

Given the artificial condition, (6.3), we can write the 
 decoherence condition as
\begin{equation}
(Tr\rho^2)^{-1} Re Tr(\rho C_{\alpha^\prime} \rho C^\dagger_\alpha) =
\delta_{\alpha^\prime \alpha} p(\alpha)\, , 
\label{sixfour}
\end{equation}
where $p(\alpha)$ is the probability of the history $\alpha$.
Summing over all the $\{\alpha_n\}$ and $\{\alpha^\prime_n\}$ except
$\alpha_k$ and $\alpha_k^\prime$, we have
\begin{equation}
(Tr\rho^2)^{-1} Re Tr [\rho P^k_{\alpha^\prime_k}(t_k) \rho
P^k_{\alpha_k}(t_k)] =
\delta_{\alpha^\prime_k \alpha_k} p(\alpha_k)\, .
\label{sixfive}
\end{equation}
We note that $P^k_{\alpha_k}(t_k)$ and $P^k_{\alpha^\prime_k}(t_k)$ are just
projection operators and thus of the form
\[
\sum_n|n><n|\quad {\rm and}\quad \sum_{n^\prime}|n^\prime><n^\prime|
\]
respectively, where the $|n>$ and $|n^\prime>$ are mutually orthogonal for
$\alpha_k\not=\alpha^\prime_k$.  Eq.~(6.5) then tells us that
\begin{equation}
(Tr\rho^2)^{-1}\ \sum_{n,n^\prime} |<n|\rho|n^\prime>|^2=0~~  {\rm for}
\ \alpha_k\not=\alpha^\prime_k\, .
\label{sixsix}
\end{equation}
Thus $\rho$ has no matrix elements between any $|n>$ and any
$|n^\prime>$ for $\alpha_k\not=\alpha^\prime_k$.  In other words, $\rho$
commutes with all the $P$'s and therefore with all the chains
$C_\alpha$ of $P$'s:
\begin{equation}
[C_\alpha, \rho] = 0~~  {\rm for~all}\ \alpha\, .
\label{sixseven}
\end{equation}
This consequence of perfect decoherence for the special case \eqref{sixthree}
has some important implications.  For one thing, the
decoherence formula can now be written
\begin{equation}
(Tr\rho^2)^{-1} Tr( C_{\alpha^\prime} \rho^2 C^\dagger_\alpha) =
\delta_{\alpha^\prime\alpha} p(\alpha)\, ,
\label{sixeight}
\end{equation}
so that we are back to ordinary quantum mechanics with only an initial
density matrix $\bar\rho\equiv (Tr\rho^2)^{-1}\rho^2$ [{\it cf.}~\eqref{fourfive} in the
classical case] but with the very
restrictive condition
\begin{equation}
[C_\alpha, \bar\rho]= 0~~ {\rm for ~~all}\ \alpha\, .
\label{sixnine}
\end{equation}
The cosmology with the symmetry (6.3) was supposed to be in contrast  
to the
usual one with only an initial density matrix, and yet it turns out to
be only a special case of the usual one with the stringent set of
conditions \eqref{sixnine} imposed in addition.  The resolution of this apparent
paradox is that Eq.~\eqref{sixnine} permits essentially no dynamics and thus
achieves symmetry between $\rho_i$ and $\rho_f$  in a rather trivial way.
That is not surprising in view of the nature of this condition discussed
above.

We have seen that any $P^k_{\alpha_k}(t_k)$ has to commute with
$\bar\rho$ if it is to be permitted in a chain of $P$'s constituting a
member of a decohering set of alternative coarse-grained histories.  Now
it is unreasonable that for a given projection operator $P$
there should be only a discrete set of times at which it is permissible
to use it in a history (e.g., for a measurement). Thus we would expect
that there should be a continuous range of such times, which means that
${\dot P} =-i[P,H]$
must commute with $\bar\rho$.  But $\bar\rho$ and ${P}$, since they
commute, are simultaneously diagonalizable, with eigenvalues $\pi_i$
and $q_i$ respectively.  The time derivative of the probability
$Tr(\bar\rho P)$ is 
\begin{equation}
Tr(\bar\rho\dot P) = -i Tr(\bar\rho[P,H]) = -i \sum_i \pi_i (q_i-q_i)
H_{ii}=0\, .
\label{sixten}
\end{equation}
The probabilities of the different projections $P$ remain constant in
time, so that there is essentially no dynamics and certainly no second
law of thermodynamics.

\subsection{Classicality}

A theory of the boundary conditions of the universe must imply the
quasiclassical domain of familiar experience.  A set of histories
describing a quasiclassical domain must, of course, decohere.  That is
the prerequisite for assigning probabilities in quantum mechanics.
But further, the probabilities must be high that these histories are
approximately correlated by classical dynamical laws, except 
for the intervention of
occasional amplified quantum fluctuations.

There are, of course, limitations on classical two-time boundary conditions.
We cannot, for example, specify both co\"ordinates and their conjugate
momenta at {\it both} an initial and a final time.  There would, in
general, be no corresponding solutions of the classical equations of
motion.
Even if initial and final conditions in quantum cosmology allow for
decoherence as discussed above, they could still be too restrictive to
allow for classical correlations.  One would expect this to be the case,
for example, if they required a narrow distribution of both
coordinates and momenta both initially and finally.  Quantum cosmologies with two 
boundary conditions are therefore limited by both
decoherence
and classicality.

\section{Conclusions}

Time-symmetric  quantum cosmologies can be constructed utilizing a
time-neutral generalized quantum mechanics of closed systems with
initial and final conditions related by time-inversion symmetry. From
the point of view of familiar quantum mechanics such time-symmetric
cosmologies are highly unusual.
If we think of Hilbert space as finite-dimensional, we could introduce a
normalization $Tr(\rho_f)\ Tr(\rho_i) = Tr(I)$, which would agree with
the usual case $Tr(\rho_i)=1$, $Tr(\rho_f) = Tr(I)$.  (Note that both
$N=Tr(\rho_f\rho_i)$ and the quantity $Tr(\rho_f)\ Tr(\rho_i)$ are
invariant under multiplication of $\rho_i$ by a factor and $\rho_f$ by
the inverse factor.)  With this normalization we may think of $N^{-1} =
Tr(\rho_f\rho_i)$ as a measure of the likelihood of the 
 final condition given the initial
one.  The similarly defined quantity $N^{-1}$ in the analogous classical
time-symmetric cosmologies is just that. It is the
fraction of trajectories meeting the
initial condition that also meet the final one [cf.~(4.1)]. The measure
$N^{-1}$ is unity for the usual cases where $\rho_f = I$.  It can be
expected to be very small for large systems with
 time-symmetric boundary conditions, 
as the simple model described in Figure 1 suggests.
The measure $N^{-1}$ is likely to be {\it extraordinarily} small in the case
of the universe itself.  Were it exactly zero the initial and final
boundary condition construction would
become doubtful.  We are unsure how much of that doubt survives if it is
merely extraordinarily small.  

As a prerequisite for a time-symmetric quantum cosmology,
the fundamental Hamiltonian 
must be time-inversion symmetric to
give a meaningful notion of time-symmetry and this restricts the
mechanisms by which the effective $CP$ violation in the weak
interactions can arise.  There must be some impurity in
the initial or final density matrices or in both for any non-trivial
probabilities to be predicted at all. If we wish to exclude   
the highly unorthodox quantum mechanics in which $|\Psi_i\rangle$ and 
$|\Psi_f\rangle$
determine all the throws of the quantum dice, then  
we could not have, for
example, a time-symmetric quantum cosmology with both the initial and final
conditions resembling something like  the ``no-boundary''
proposal. These
results have been obtained by assuming unrealistic
exact decoherence and by neglecting gross quantum variations in the
structure of spacetime, which may be important in the early universe.  
  It would be desirable to extend the discussion to remove
these special restrictions.  

Even if these purely theoretical requirements for time-symmetry were met,
observations might rule out such boundary conditions.  Deviations from the
usual thermodynamic or $CP$ arrows of time may be undetectably small if
the time between initial and final conditions is long enough.  But, as
suggested by Davies and Twamley, an expanding and contracting
time-symmetric cosmology may be transparent enough to electromagnetic
and other forms of radiation that the effects of  time-symmetric initial
and final
conditions would be inconsistent with observations today.
In the absence of some
compelling theoretical principle mandating time symmetry,  the simplest
possibility seems to be the usually postulated 
universe where there is a fundamental
distinction between past and future --- a universe with a special
initial state and a final condition of indifference with respect to
state.  Nevertheless, the notion of complete $T$ symmetry or
$CPT$ symmetry remains sufficiently intriguing to warrant further
investigation of how such a symmetry could occur or what observations
could rule it out.  In this paper we have provided a quantum-mechanical
framework for such investigations. 

\acknowledgments

An earlier version of this paper appeared in the {\sl
Proceedings of the 1st International Sakharov Conference on Physics},
Moscow, USSR, May 27--31, 1991 as a tribute to the memory of
A.D.~Sakharov.

Part of this research was carried out at the Aspen Center for Physics.
The work of MG-M was supported by DOE contract DE-AC-03-81ER40050 and by
the Alfred P. Sloan Foundation.  That of JBH was supported by NSF
grant PHY90-08502.

\section*{Appendix}

We derive the expression (5.12) for the expected value of a scalar field
in the time-neutral quantum mechanics of cosmology with an initial
condition represented by a density matrix $\rho_i$ and a final condition
represented by a density matrix $\rho_f$.  Consider alternatives such that the
value of the field $\phi$ at $(\vec x, t)$ lies in one of an exhaustive
set of infinitesimal exclusive intervals $\{\Delta_\alpha\}$
 with central values
$\{\phi_\alpha\}$.  Let $P_\alpha(\vec x, t)$ denote the corresponding
projection operators.  The decoherence functional for this set of
alternatives is, according to (6.1)
\begin{equation}
D(\alpha^\prime, \alpha) = NTr\left[\rho_f P_{\alpha^\prime} (\vec x, t)
\rho_i P_\alpha (\vec x, t)\right]\, .
\label{appone}
\end{equation}
We assume that this is diagonal, that is, proportional to
$\delta_{\alpha\alpha^\prime}$. The diagonal elements give the
probabilities of the alternative values of the field according to (3.1).
Thus the expected value of $\phi(\vec x, t)$ is
\begin{equation}
\langle\phi (\vec x, t)\rangle = \sum\nolimits_\alpha \phi_\alpha
D(\alpha, \alpha)\, .
\label{apptwo}
\end{equation}
Because the alternatives decohere, this can be written in two equivalent
forms
\begin{equation}
\langle \phi (\vec x, t) \rangle = \sum\nolimits_{\alpha^\prime\alpha}
\phi_{\alpha^\prime} D(\alpha^\prime, \alpha) =
\sum\nolimits_{\alpha^\prime\alpha} \phi_\alpha D(\alpha^\prime, \alpha)
\, .\label{appthree}
\end{equation}
But,  utilizing $\sum_\alpha P_\alpha (\vec x, t) = 1$ and $\sum_\alpha
\phi_\alpha P_\alpha (\vec x, t)= \phi (\vec x, t)$, as well as (A.1) and the
cyclic property of the trace, we get
\begin{equation}
\langle \phi (\vec x, t) \rangle = NTr \left[\rho_f \phi (\vec x, t)
\rho_i\right] = NTr\left[\rho_i \phi (\vec x, t) \rho_f\right]
\label{appfour}
\end{equation}
as in \eqref{fivetwelve}.

\section*{Questions and Answers}

B.~DeWitt: If you propose that the universe is in a particular quantum
state, determined by particular initial conditions, why do you bother
with a complete Hilbert-space framework for discussion?
\vskip .13 in
JH: I interpret your question as mainly referring to the status of the
superposition principle in quantum cosmology.  It is true that if the
initial condition of our universe is described by a single wave function
then it is never necessary to superpose it with another to make
predictions.
  However, the principle of superposition enters centrally
elsewhere in the predictive framework.  Specifically, it enters into the
construction of the probabilities for the coarse-grained sets of
histories that we observe.  If $P_A$ and $P_B$ are projection operators
  representing exclusive
alternatives $A$ and $B$ at one time, then the alternative $A$ {\it and}
$B$ is represented by the {\it sum} of the projections, $P_A + P_B$.
That is a specific instance of the principle of superposition.  More
generally, the decoherence functionals for sets of histories related by
an operation of coarse graining must be connected by the superposition
principle.  That is one reason we assume the full apparatus of Hilbert
space when discussing the quantum cosmology of matter fields in a fixed
background spacetimes or generalizations of that formalism consistent
with the superposition principle when quantum gravity is taken into
account. Even in the most general cosmological context it should still
be possible to test these aspects of the principle of superposition.
\vskip .26 in
K.~Kucha\v r: Both Murray's and your talk were based on the assumption
that there is a true Hamiltonian and that there is a single time
parameter which orders the projection operators. How does one formulate
the difference between time-symmetric initial and final conditions if
the dynamics is driven by constraints and there is no privileged time
parameter?
\vskip .13 in
JH: To keep the discussion in the talk manageable, we assumed a fixed,
background spacetime. That, of course, is an excellent
approximation anytime much more than a Planck time after the initial
singularity and a Planck time  before the final singularity if there is
one.
That fixed spacetime geometry supplies the notion of time used to
order the operators and define the Hamiltonian.
However, in regimes near the singularity, where quantum gravity is
important and the geometry of spacetime fluctuates quantum
mechanically, there will be no fixed spacetime geometry to supply
a notion of time. A further generalization of quantum mechanics is
thus required. I have described in several places the basic elements
of one such generalization
based on sum-over-histories quantum mechanics.\footnote{See, for example,
my lectures ``The Quantum Mechanics of Cosmology''
 in  {\sl Quantum
Mechanics and Baby Universes: Proceedings of the 1989 Jerusalem Winter
School}, edited by S. Coleman, J. Hartle, T. Piran, and S. Weinberg,
World Scientific, Singapore, 1990, or in more complete detail, in my
lectures ``Spacetime Quantum Mechanics and the Quantum Mechanics of
Spacetime'' in {\sl Gravitation and Quantizations}, Proceedings of the
1992 Les Houches Summer School, ed.~by B.~Julia and J.~Zinn-Justin, North
Holland, Amsterdam, 1993.} In that generalization, the
histories
are four dimensional cosmological spacetimes with boundaries where the
analogs of ``initial'' and ``final'' conditions represented
by wave functions are imposed. The decoherence functional for
coarse-grained sets of alternative histories of the universe, including
diffeomorphism invariant coarse grainings of spacetime geometry, is
represented in a sum-over-histories form that does not single out a
privileged time parameter.  Decoherence is thus defined and
probabilities for the individual members of decoherent sets of
coarse-grained histories can be calculated.
In order not to be manifestly inconsistent
with observations, the specific initial and final conditions of our
universe had better predict the approximately classical behavior of
spacetime geometry on accessible scales in our epoch. That is,
semiclassically, realistic boundary conditions predict
 an ensemble of possible classical spacetimes
of which we live in one.  The probabilities of suitably
coarse-grained matter field histories in each spacetime in the ensemble
would be approximately given by the kind of quantum mechanics we have
limited ourselves to in this talk with a notion of time given by the
particular classical spacetime geometry.  The discussion we gave is thus
both a model for the more general case of quantum gravity and an
approximation to it in all directly accessible circumstances.

If the initial and final conditions are suitably related, I would expect
the ensemble of possible spacetimes predicted semiclassically by such a
theory to exhibit statistical time symmetry.  Further, for the probable
spacetimes that are time symmetric I would expect the quantum mechanics
of matter fields and small fluctuations of geometry to be time-symmetric
in the sense have described in this talk.  Put briefly, I
expect the present discussion that assumes a fixed spacetime is a good
approximation in interesting circumstances to the more general case
where it is allowed to fluctuate.
  It is fair to say, however, that detailed
calculations have not been done to check on these expectations.
\vskip .26 in
R.~Omn\'es: Your probability formula, in a universe with a destiny,
violates Gleason's theorem and therefore one of its assumptions at
least. There are two possibilities: (i) Not all properties are possible,
which is what you are aiming at. (ii) Maybe, the Hilbert space is
highly non-separable.
\vskip .13 in
JH: I think it's the former.
\vskip .26 in
J.~Halliwell: You argue that for initial and final density matrices
satisfying a certain condition (and in particular, for pure initial and
final states) the decoherence functional factors, and therefore, will
not decohere except for certain trivial histories.  You suggested that
there will therefore be problems for the no-boundary state, which is a
pure state.  It seems to me that this result may depend rather crucially
on the existence of a Hilbert space structure etc., and in particular,
on the possibility of folding in initial and final states using the
usual inner product.  My point is that all of this structure is not
known to exist for quantum cosmology.  The decoherence functional for
quantum cosmology is yet to be constructed, and is likely to have a
structure rather different to the quantum mechanical one.  You may
therefore be premature in your conclusions about the no-boundary
proposal.
\vskip .13 in
JH: In the proposals for the decoherence functional
for quantum cosmology that I have put forward, the result
that pure initial and final states permit the decoherence of only
trivial sets of histories continues to hold in much the same way it does
for the quantum cosmologies in a box described in the text.  Initial and
final conditions are represented by density matrices and pure conditions
by single wave functions.  The principle of superposition of amplitudes
is maintained as is the relation between amplitudes and probabilities.
When the initial and final states are pure the decoherence functional
factors into a term for one history times a term for the other as in
(6.2) and the rest of the argument goes through.  In these
generalizations imposing the ``no boundary'' proposal for both initial
and final conditions does not lead to interesting sets of decohering
histories.  However, there is much to be investigated here and there
could be other generalizations of quantum mechanics for which the result
does not hold.
\vskip .26 in
I.~Bialynicki-Birula: I would like to make a comment on the possible
role of soft photons in the time-symmetric quantum theory.  In the
presence of massless particles, and that is a typical case, even for
time-symmetric Hamiltonians there is a difficulty in implementing the
time-symmetry condition,
\[
\rho_f = {\cal T}^{-1} \rho_i {\cal T}\, ,
\]
due to the existence of infrared radiation.  There is no $\rho_f$ that
will give a nonvanishing transition probability whenever charged
particles are being accelerated.  In order to obtain a finite result we
must perform an integration over the momenta of final, unobservable soft
photons.  The necessity to perform this integration implies that there
is an asymmetry between the initial state described by $\rho_i$ and the
final state for which a density operator does not exist.
\vskip .13 in
JH: Soft photons in the universe certainly provide an important and
widespread mechanism for decoherence.  However, I haven't thought
through the effect soft photons might have on the size of the
normalizing factor $N^{-1} = Tr (\rho_f\rho_i)$ that occurs in the
expression (3.1) for probabilities.  I would be surprised if, in a
proper formulation of quantum electrodynamics, $N^{-1}$ {\it
necessarily} vanished identically for the finite (although
cosmologically long) time interval between time-symmetric initial and
final conditions that we have been discussing.  However, even if
$N^{-1}$ does not vanish identically, but is only very small, that would
signal a significant difference between the statistics of histories in a
time-symmetric universe and the usual case.  One guesses that $N^{-1}$
is likely to be small in any realistic time-symmetric even in the
absence of electrodynamics but if soft photons play a significant role
in determining its size that would be very interesting. 
\vskip .26 in
D.~Page: What are the consequences if you require a $CPT$-invariant
universe instead of a T-invariant universe?
\vskip .13 in
JH:  I didn't get to $CPT$-symmetric cosmologies in my talk, but they
are discussed in the written contribution that Murray Gell-Mann and I
have submitted to the proceedings. In the generalization of quantum
mechanics that we discuss, a $CPT$-symmetric universe will
result if the initial and final density matrices are related by a $CPT$
transformation. Since all local field theories are $CPT$-invariant,
$CPT$-symmetric cosmologies are possible even if the $CP$ violation observed
in the weak interactions arises from a fundamental Hamiltonian that is
$CP$-non-invariant. That is in contrast to the case of T symmetry which
can only be achieved with a $CP$-symmetric Hamiltonian so that the
observed $CP$ violation must arise from one of the symmetry breaking
mechanisms we discussed. However, also as discussed in the
text, $CPT$ symmetry may be difficult to reconcile with universes that
are homogeneously dominated by matter near one singularity (and
therefore antimatter dominated near the other) unless the lifetime
is very long.
\vskip .26 in
B.~DeWitt: So if the no boundary condition leads to a $CP$ invariant
family of $CP$ violating states, you must pick one member of this family
out by hand?
\vskip .13 in
JH: If the ``no boundary'' condition leads to a $CP$ invariant family of
$CP$ non-invariant states then I would prefer to say that the initial
condition is a density matrix with probabilities distributed uniformly
among these possibilities.  But for predictive purposes that amounts to
what you said.
\vskip .26 in
A.~Albrecht: If one discusses the thermodynamic arrow of time in a time
symmetric universe, one has a region near the ``beginning'' where the
arrow
runs toward the middle, and a region near the ``end'' where the arrow
runs towards the middle. In the middle there is no particular arrow of
time. The probability that we survive into the future epoch where the
arrow is reversed is no greater than the probability that some IGUS
is present right now, evolving in the opposite direction of time.
\vskip .13 in
JH:
I think that's essentially right with a few qualifications. It follows
from the assumed time symmetry that the statistics of IGUS's at the
present age from the big bang must be the same as that at a comparable
time from the final singularity if the only input to estimating those
statistics  is the initial
and final conditions.  If the lifetime of the universe is long compared
to those times both the probability that an IGUS survives into the far
future
and the probability
 that there are IGUSes in the present living backward in time may
be very low.  If we are a typical IGUS then those probabilities
apply to us.  However, we have more information about our particular
history with which the conditional probabilities of our surviving into
the far future could, in principle, be calculated and compared with the
probability that there are IGUSes that have evolved backward from the
far future around today.  I'll leave it to you to make the estimate of
whether our particular history makes it more or less probable that we
survive farther into the future than the typical IGUS!
\vskip .26 in
P.~Davies: Your model is most plausible if all asymmetric physical
processes relax to equilibrium before the time reversal occurs. But long
ago it was found that the future light cone in Robertson-Walker
cosmological models is transparent to photons. (In the recontracting
phase it is transparent at least until the turnover point.) Thus,
retarded radiation cannot equilibrate before time reversal, so that the
imposing of time-symmetric boundary conditions would surely show up
experimentally in the emission of radiation.
\vskip .13 in
JH: As mentioned in the talk, there are several different examples of
physical systems that will not come to equilibrium in the Hubble time,
and the system of matter and radiation is one of them.  I take
your comment to be a suggestion that observation in the electromagnetic
system could supply the best lower bound on the time between initial and
final conditions beyond which these are indistinguishable.
That may well be the case and is an interesting
subject for further research.  It's an important question. [For further
discussion see Section IVc, added after the conference.]
\vskip .26 in
A.~Starobinsky: (in response to P.~Davies' comment)\hfill\break
This is not always the case.  If the width of a domain wall in flat
spacetime is larger than the radius of curvatures ``gravitational
radius'' corresponding to the field energy density in a metastable state
at the top of the field potential, then regions with the different sign
of $CP$ violation are always beyond an observer's particle horizon and
they are never accessible to him for any future evolution.  A good
example can be constructed using the ``new'' inflationary scenario and
relating the sign of an inflation field to the sign of $CP$ violation.

\section*{Note added by M~Gell-Mann and J.B.~Hartle, November 1993.}

We would like to clear up any possible confusion over the relationship among several ideas in
the quantum mechanics of closed systems as well as the history of the subject and the
terminology employed.  In this note, added two years after the discussion, we attempt to clarify
these matter as we see them, benefiting from research in the intervening time.

As far as we are aware, in the context of the modern interpretation of quantum mechanics, the
term ``decoherence'' was first used by us in lectures and discussion after 1986 (summarized in
our paper published in 1990) to describe a property of a set of alternative time-{\sl histories} of
a closed system.  Specifically a set of histories is said to (medium) decohere when there is
negligible or vanishing quantum mechanical interference between the individual histories in the
set as measured by the ``off-diagonal'' elements of the decoherence functional $D(\alpha^\prime,
\alpha)$.  However, the term ``decoherence'' has subsequently come to be used to refer also to
the decay over time of the off-diagonal elements of a reduced density matrix defined by a
division of a complete set of variables into ones that are traced over in constructing the
reduced density matrix and the rest.  The decay was discussed in connection with the
interpretation of quantum mechanics in the '70s and early '80s by Zeh, Zurek, and others
(although not referred to as decoherence) and by many others since.  These two notions of
decoherence --- of histories and of density matrices --- are not the same but are not
unconnected either.  The vanishing at a sequence of times of the off-diagonal elements of a
reduced density matrix in certain variables is neither mathematically nor physically equivalent
to the decoherence of the corresponding set of alternative histories.  However, the ideas are
connected in certain idealized models where it can be shown that the physical mechanisms
causing the decoherence of histories coarse-grained by ranges of values of certain sets of
coordinates suitably spaced in time, {\it also} lead to the diagonalization of the reduced
density matrix in these variables over similar intervals of time \cite{Har91}.  It has also been shown
that in similar models, under restrictive conditions, the diagonalization of the reduced density
matrix implies the decoherence of the histories associated with the ``Schmidt basis'' for that
density matrix at suitable times \cite{PZ93, GH93a}. Finally, a certain interpretation of ``mechanism of
decoherence'' can be defined \cite{Fin93, GH95} that generalizes the reduced density matrix concept of
decoherence {\it in the context of a stronger form of decoherence of histories}. A precise
connection is thereby established between the two kinds of decoherence.  The reader should
therefore keep in mind that in these and other discussions and papers at this conference the
word ``decoherence'' is used for two distinct but connected ideas --- the decoherence of reduced
density matrices and the decoherence of histories.

We now try to clarify the connection of our ideas with the work of Bob Griffiths and Roland
Omn\` es, another topic that was raised in the discussion.  In any quantum mechanical theory, a
rule is needed to discriminate between those sets of histories that can be assigned
probabilities and those that cannot because of quantum mechanical interference. Griffiths was
the first to propose a quantum mechanics of closed systems with a rule that did not involve a
fundamental notion of measurement.  Instead, probabilities are assigned to just those sets of
histories that are ``consistent'' in the sense that their probabilities obey correct sum rules
--- essentially  the consistency requirement that the theory not offer two different results for
the same probability. (In this connection, it may be helpful to note that what is sometimes
referred to as the quantum mechanics of an ``open system'' means a set of effective rules for
describing the quantum mechanical behavior of {\it part} of a closed system.)  Griffiths' ideas
were extended by Omn\` es and a similar formulation was arrived at independently, but later, by
ourselves.  As far as we are aware, {\it all} formulations of the quantum mechanics of closed
systems under serious consideration, including ours, are ``consistent history formulations'' in
the sense of requiring the consistency of a set of probability sum rules for those histories
that are assigned probabilities.  However, there are different formulations depending on just
what probability sum rules are required and the strength of the conditions used to ensure their
consistency.

The consistency conditions of Griffiths are not the same as the decoherence conditions used in
our work.  To explain the difference, a small amount of notation is useful. Let $\{C_\alpha\}$
denote a set of chains of projections representing a set of alternative histories, and
$D(\alpha^\prime, \alpha) = Tr(C_{\alpha^\prime} \rho\, C_\alpha)$ the associated decoherence
functional.  The consistency condition of Griffiths is that $Re\, D(\alpha^\prime, \alpha)
\approx 0$, $\alpha^\prime \not= \alpha$, if and only if $C_{\alpha^\prime} + C_\alpha$ is {\it
another chain of projections}. This is the necessary condition for the probability sun rules if
only histories corresponding to independent sets of alternatives at different moments of time
(histories which are represented by chains of projections) are allowed.  What we called the
weak decoherence condition, $Re\, D(\alpha^\prime, \alpha) \approx 0$, $\alpha^\prime \not=
\alpha$, is {\it stronger} than that of Griffiths.  It is the necessary condition if histories
that are {\it sums} of chains of projections are allowed (corresponding to a rule for
coarse-graining that allows arbitrary unions of histories as new histories).  Our medium
decoherence condition, $D(\alpha^\prime, \alpha) \approx 0$, $\alpha^\prime \not= \alpha$, is
{\it still stronger}, implying both of the above conditions but not being implied by either of
them.  Thus, there are at least three different notions of decoherence of histories, {\it all}
of which imply the consistency of (sometimes different) sets of probability sum rules.

In our work we have sought a notion of decoherence of histories that would capture generally the
idea that, in physically interesting situations, relevant for quasiclassical behavior, quantum
mechanical interference between histories vanishes {\it for a reason}.  That is, we sought to
provide a general characterization of the mechanisms of dissipation of phases described by Zeh,
Zurek, and others, but in a way that would not require an artificial or poorly defined division
of the closed system into subsystem and environment.  For this reason we were led to notions of
the decoherence of histories that imply consistency, of course, but are stronger and should not
be confused with it.

The differences we have described should not obscure the fact that our work lies within the
class of consistent histories formulations of the quantum mechanics of closed systems, but
should also not obscure the fact that decoherence of histories, as we have defined it, is not
the same as just consistency.

\end{document}